\documentclass[11pt]{article}
\usepackage{jheppub}

\usepackage{color}
\usepackage{colordvi}
\usepackage{changes}
\usepackage{epsfig,epsf}
\usepackage{amsmath}
\usepackage{amsthm}
\usepackage{amsfonts}
\usepackage{amssymb}
\usepackage{dsfont}
\usepackage[mathscr]{euscript}
\usepackage{eufrak}
\usepackage{epstopdf}
\usepackage{multirow}
\usepackage{mathrsfs}
\usepackage{rotating}

\usepackage{marvosym}

\usepackage{todonotes}

\usepackage{subcaption}
\usepackage{array}   % for \newcolumntype macro
\newcolumntype{C}{>{$}c<{$}}

\usepackage{slashed}

\usepackage[active]{srcltx}
\newcommand{\dhalf}{\frac{d}{2}}

\newcommand{\Y}{Y}

\newcommand{\DIS}{\mathrm{\scriptscriptstyle DIS}}

\def\II{\hbox{{1}\kern-.25em\hbox{l}}}

\DeclareMathOperator{\Li}{Li}

\DeclareMathOperator{\GPL}{G}

\def\II{\hbox{{1}\kern-.25em\hbox{l}}}

\makeatletter
\renewcommand\@fpheader{}
\renewcommand\@journal{}
\makeatother

\title{
\begin{flushright}
{\large \textnormal{DESY 24-178}}\\[2mm]
\end{flushright}
The two-loop coefficient functions for double deeply virtual Compton scattering}

\author[a]{Vladimir M. Braun,}
\author[b]{Hua-Yu Jiang,}
\author[c]{Alexander N. Manashov,}
\author[a]{and Andreas~von~Manteuffel}

\affiliation[a]{
   Institut f\"ur Theoretische Physik, Universit\"at
   Regensburg, D-93040 Regensburg, Germany}
\affiliation[b]{
Institute of Particle and Nuclear Physics, Henan Normal University, Xinxiang 453007, Henan, P. R. China}
\affiliation[c]{
   II.~Institut f\"ur Theoretische Physik, Universit\"at Hamburg,
   D-22761 Hamburg, Germany}

\emailAdd{vladimir.braun@ur.de}

\emailAdd{jianghuayu@htu.edu.cn}

\emailAdd{alexander.manashov@desy.de}

\emailAdd{manteuffel@ur.de}

\abstract{
Making use of conformal symmetry of large-$n_f$ QCD in $d=4-2\epsilon$ dimensions 
at the Wilson-Fischer fixed point, we calculate the two-loop coefficient 
functions in the operator product expansion of two electromagnetic currents in general kinematics
with two different photon virtualities. This result is necessary for the description of
the double deeply virtual Compton scattering to the next-to-next-to-leading order accuracy,
but is also interesting for a range of other two-photon processes.
We present analytic expression for the coefficient function in momentum fraction space in the
$\overline{\text{MS}}$ scheme and study its numerical impact on the
Compton form factors for a simple model of the generalized parton distributions.
The calculated corrections turn out to be large and are significant for the kinematics of 
proposed experiments. 
       }


\begin{document}
\maketitle

%%%%%%%%%%%%%%%%%%%%%%%%%%%%%%%%%%%%%%%%%%%%%%%%%%%%%%%%%%%%%%%%%%%%%%%%%%%%%%%%%%%%%%%%%%%%%%%%%%%%%%%%%%%%%%%%%%%%%%

                          \section{Introduction}\label{sec:intro}

%%%%%%%%%%%%%%%%%%%%%%%%%%%%%%%%%%%%%%%%%%%%%%%%%%%%%%%%%%%%%%%%%%%%%%%%%%%%%%%%%%%%%%%%%%%%%%%%%%%%%%%%%%%%%%%%%%%%%%

Deeply-virtual Compton scattering (DVCS) \cite{Ji:1996nm,Radyushkin:1996nd,Mueller:1998fv} is generally accepted to be the
``gold-plated'' process with the highest potential impact on the determination of the generalized parton distributions (GPDs)
in the nucleon. The problem is, however, that at leading order (LO) the DVCS and time-like Compton scattering (TCS) amplitudes
only involve GPDs at the $x=\xi$ line, where $x$ is the average parton momentum and $\xi$ is the asymmetry parameter. The
double deeply virtual Compton scattering $\gamma^\ast(q_1) + N(p_1) \to \gamma^\ast(q_2) + N(p_2)$ (DDVCS) avoids this restriction
\cite{Belitsky:2002tf,Guidal:2002kt} and can be accessed by studying exclusive electroproduction of a lepton pair.
Varying the invariant mass of the lepton pair, one can, in principle, directly extract the GPDs from the observables.
DDVCS can be measured in near future at both fixed target \cite{Zhao:2021zsm} and collider facilities \cite{AbdulKhalek:2021gbh,Anderle:2021wcy}.
A preliminary impact study of DDVCS phenomenology for the JLAB12, JLAB20+ and EIC kinematics \cite{Deja:2023ahc}
reached promising conclusions.

The main challenge of all GPD studies is that the quantities of interest are functions of three kinematic variables.
Their extraction requires a massive amount of data and very high precision for both experimental and theory inputs.
The future GPD determinations will therefore have to be based on global fits of all available experiments and the constraints from lattice measurements
and PDFs in the forward limit.
It is imperative that all ingredients in such fits are calculated with the same precision.
Ideally, one would like to reach the same level of accuracy as in inclusive reactions, where the next-to-next-to leading
order (NNLO)  analysis has become the standard in the field~\cite{Accardi:2016ndt}.
One-loop DVCS coefficient functions have been known for a long time \cite{Ji:1997nk,Belitsky:1997rh} and 
the two-loop ones have been calculated recently~\cite{Braun:2020yib,Gao:2021iqq,Braun:2021grd,Braun:2022bpn,Ji:2023xzk}.
Two-loop evolution equations for the GPDs are known from \cite{Belitsky:1998gc,Belitsky:1999hf}.
Three-loop evolution equations for flavor-nonsinglet GPDs in position space have been derived in \cite{Braun:2017cih,Ji:2023eni}
and for the first few moments of flavor-singlet GPDs in \cite{Braun:2022byg}. The DDVCS description has to be extended to the same level of accuracy.
As the first step in this direction, in this work we calculate the two-loop DDVCS coefficient functions (CFs) for the flavor-nonsinglet vector contributions
using conformal symmetry techniques.

The idea to apply the conformal symmetry to off-forward reactions  is not new
but the early work \cite{Brodsky:1984xk} was missing an important element: the scheme-dependent difference
between the dilatation and special conformal anomalies~\cite{Mueller:1991gd}.
It was first shown in~\cite{Belitsky:1997rh} that conformal symmetry 
provides a connection between the CFs in DVCS and DIS.
The general strategy of our calculation follows Ref.~\cite{Braun:2020yib}, but involves some new technical elements.
We make use of conformal symmetry of large-$n_f$ QCD in non-integer $d=4-2\epsilon$ dimensions at 
the Wilson-Fischer fixed point~\cite{Braun:2013tva,Braun:2018mxm}. In a conformal theory the contributions of operators with total derivatives
are related to the contributions of the operators without total derivatives by symmetry transformations and do not need to be calculated
separately. In this way, the calculation of the  $\ell$-loop off-forward CF
can be reduced to the  $\ell$-loop forward CF, known from DIS, and the $(\ell-1)$-loop calculation of the off-forward CF in
$4-2\epsilon$ dimensions, including terms $\mathcal{O}(\epsilon^{\ell-1})$.

The presentation is organized as follows.
Section~\ref{sec:general} is introductory, it contains
general definitions and specifies our notation and conventions.
In Sect.~\ref{sec:framework} we present the general framework and the procedure for the calculation
of CFs in the OPE of two electromagnetic currents using conformal
symmetry of QCD at the Wilson-Fischer fixed point in non-integer dimensions.
A new ansatz for the solution is presented, which allows one to solve the relevant equations for the case of arbitrary photon 
virtualities. 
Section~\ref{sec:twoloopCFs} is devoted to the particularities of the two-loop calculation 
and the discussion of the mathematical structure of the results. 
Explicit expressions for the CFs in momentum fraction space are
presented in Appendix~\ref{App:TwoLoopCFs} and in two ancillary files using different representations for the 
relevant generalized polylogarithms. Numerical estimates of the size of the two-loop correction for realistic 
kinematics are presented in Sect.~\ref{sec:numerics}. The final Sect.~\ref{sec:summary} is reserved for 
a short summary. The paper also contains several Appendices explaining the construction of helicity amplitudes, some useful integrals, expansion 
of the CFs in the threshold region, and more.

%%%%%%%%%%%%%%%%%%%%%%%%%%%%%%%%%%%%%%%%%%%%%%%%%%%%%%%%%%%%%%%%%%%%%%%%%%%%%%%%%%%%%%%%%%%%%%%%%%%%%%%%%%%%%%%%%%%%%%

               \section{Kinematics, notation and conventions, and one-loop results}\label{sec:general}

%%%%%%%%%%%%%%%%%%%%%%%%%%%%%%%%%%%%%%%%%%%%%%%%%%%%%%%%%%%%%%%%%%%%%%%%%%%%%%%%%%%%%%%%%%%%%%%%%%%%%%%%%%%%%%%%%%%%%%

The generalized Compton amplitude  is given by a Fourier transform of the off-forward matrix element of the time-ordered product 
of two electromagnetic currents,
\begin{align}
A_{\mu\nu} & = i \int d^4 x \, e^{i q_1\cdot x}
	\big\langle p_2 \big\lvert T\big\{ j_\mu^{\mathrm{em}}(x) \, j_\nu^{\mathrm{em}}(0) \big\} \big\rvert p_1 \big\rangle,
\label{Compton}
\end{align}
corresponding to the double deeply virtual Compton scattering process
\begin{equation}
\gamma^\ast(q_1) + N(p_1) \rightarrow \gamma^\ast(q_2) + N(p_2)\,.
\end{equation}
Here $q_1$ and $q_2$ are the momenta of the incoming and outgoing photons, respectively, $p_1,p_2$ are the target (nucleon) momenta
in initial and final states, and $q_2=p_1+p_2-q_1$.
Let
\begin{align}
& q = \frac12(q_1 + q_2), \qquad p = \frac12 (p_1 + p_2), \qquad \Delta = p_2 - p_1 = q_1 - q_2\,, \qquad Q^2 = - q^2,
\end{align}
and
\begin{align}\label{xi-eta-omega}
\xi = - \frac{\Delta\cdot q}{2 p\cdot q}, \qquad \eta = \frac{Q^2}{ 2 p\cdot q},
	\qquad \omega =  \frac{\xi}{\eta}  = \frac{q_1^2 - q_2^2}{q_1^2 + q_2^2 - \Delta^2/2} .
\end{align}
In the following we assume $\Delta^2=0$.

The DVCS corresponds to $\omega = 1$ such that $\eta = \xi \simeq x_B/(2-x_B)$, DIS corresponds to $\xi=0$, $\eta = x_B$,
TCS corresponds to $\omega=-1$, and exclusive electroproduction of a lepton pair (DDVCS)  to $\omega < -1$.
For all processes of interest $q_1^2-q_2^2 < 0$ and $0<\xi<1$.

In the leading-twist approximation, the parity-even (vector) part of the DDVCS amplitude
can be written in terms of two Compton form factors (CFFs), e.g. \cite{Belitsky:2005qn}
\begin{align}
A^{\mu\nu} & =
 \left(- g^{\mu\nu} + \frac{q_2^\mu q_1^\nu}{(q_1\cdot q_2)}\right) \mathcal{F}_1(\xi,\eta, \Delta^2,Q^2)
+ \frac{2}{pq} \left( p^\mu + \frac{1}{2\eta} q_2^\mu\right) \left( p^\nu + \frac{1}{2\eta} q_1^\nu\right)
\mathcal{F}_2(\xi,\eta, \Delta^2,Q^2).
\label{F12L}
\end{align}
A more convenient decomposition is in terms of  the ``transverse'' and ``longitudinal'' CFFs defined as
\begin{equation}
  \mathcal{F}_\perp  = \mathcal{F}_1\,,
\qquad \quad
 \mathcal{F}_L  = \frac{1}{\eta} \mathcal{F}_2 - \mathcal{F}_1\,.
\end{equation}
For completeness, in Appendix~\ref{Appendix:A}, we also discuss the decomposition of the amplitude $A^{\mu\nu}$
in terms of helicity amplitudes.

The factorization theorem~\cite{Radyushkin:1997ki,Ji:1998xh,Collins:1998be} relates flavor-nonsinglet
contributions to the CFFs $\mathcal{F}_i$, $i=\perp,L$, to charge conjugation $C=+1$ combinations of quark GPDs
\begin{align}
 \mathcal{F}_i(\xi,\eta, \Delta^2,Q^2) &= \sum_q e_q^2 \int_{-1}^1\!\frac{dx}{\xi} \, C_i\left(\frac{x}{\eta},\frac{\xi}{\eta},
 \frac{Q^2}{\mu^2}\right) F_q^{(+)}(x,\xi,\Delta^2,\mu^2)\,,
\notag\\
  F_q^{(+)}(x,\xi,\Delta^2,\mu^2) &=  F_q(x,\xi,\Delta^2,\mu^2) - F_q(-x,\xi,\Delta^2,\mu^2)\,.
\label{CFFfactorization}
\end{align}
The GPDs are defined by an appropriate matrix element of the light-ray operator
\begin{align}\label{LR-operator}
\mathcal O_q(z_1,z_2)=\bar q(z_1 n)\slashed{n}[z_1n,z_2n]q(z_2n)\,,
\end{align}
where $n^\mu$ is a light-like vector, $[z_1n,z_2n]$ is the Wilson line. 
For our present purposes one can choose (see Appendix~\ref{Appendix:A}) $n^\mu = q_1^\mu/q_1^2 - q_2^\mu/q_2^2$ so that
$\xi= - \Delta_+/(2p_+)$ where $\Delta_+= \Delta\cdot n$ and $p_+= p\cdot n$.
%For brevity we we do not show the dependence on the factorization scale $\mu^2$. 
An off-forward matrix element of the renormalized light-ray operator \eqref{LR-operator} can be parametrized as follows:
\begin{align}\label{H-def}
\langle p_2|[\mathcal O_q(z_1,z_2)]|p_1\rangle & = 2p_+\int_{-1}^1 dx\,  e^{-ip_+\xi(z_1+z_2) +i p_+x (z_1-z_2)} \, F_q(x,\xi,\Delta^2,\mu^2)\,,
\notag\\
  F_q^{(+)}(x,\xi,\Delta^2,\mu^2) &=  F_q(x,\xi,\Delta^2,\mu^2) - F_q(-x,\xi,\Delta^2,\mu^2)\,.
\end{align}
where the bracket $[\ldots]$ in the matrix element above stands for renormalization in the $\overline{\mathrm{MS}}$ scheme.
The quark GPD for a nucleon can further be decomposed in contributions of the two Dirac structures
$H_q(x,\xi)$ and $E_q(x,\xi)$\cite{Diehl:2003ny},
but this decomposition is irrelevant for our present purposes.

The scale dependence of the GPDs is governed by the renormalization group equation (RGE)
\begin{align}
 \left(\mu\frac{\partial}{\partial \mu}  + \beta(\alpha_s)\frac{\partial}{\partial \alpha_s} + \mathbb{H} \right)[\mathcal O_q(z_1,z_2)] =0\,,
\label{RGE}
\end{align}
where $\mathbb{H}$ (evolution kernel) is an integral operator acting on the coordinates $z_1, z_2$.
Translation-invariant polynomials $z_{12}^N = (z_1-z_2)^N$ are eigenfunctions of the evolution
kernel and the corresponding eigenvalues define the anomalous dimensions of local operators with spin $N$,  
\begin{align}
  \mathbb{H}\, z_{12}^{N-1} = \gamma_N  z_{12}^{N-1},
\end{align}
see \cite{Braun:2017cih,Braun:2020yib} for the systematic presentation and details.

In what follows we drop electromagnetic charges and the sum over flavors, and introduce a notation
\begin{align}
    z = x/\eta\,, \qquad\quad  C_i\left(\frac{x}{\eta},\frac{\xi}{\eta}, \frac{Q^2}{\mu^2}\right)
\equiv  C_i\left(z,\omega, \frac{Q^2}{\mu^2}\right)\,.
\end{align}
The CFs do not depend on the target and can be calculated in perturbation theory
\begin{align}
C_i(z,\omega, Q^2/\mu^2 ) = C^{(0)}_i(z) + a_s C^{(1)}_i(z,\omega, Q^2/\mu^2) + a_s^2 C^{(2)}_i(z,\omega, Q^2/\mu^2) + \ldots,
\end{align}
where
\begin{align}
 a_s = \alpha_s(\mu)/(4\pi)\,.
\end{align}
They are real functions in the Euclidean region $\omega \in [-1,1]$, $z\in (-1,1)$ and $Q^2>0$, and can be continued analytically
\cite{Mueller:2012sma} to the physical regions of different processes. Assuming $x$, $\xi>0$ and $q_1^2-q_2^2 <0$ are real numbers and
the usual causal prescription for $q_1^2 + q_2^2 \mapsto q_1^2 + q_2^2+i 0$ one obtains
   \begin{align}\label{AnalyticContinuation}
  C_i\left(\frac{x}{\eta},\frac{\xi}{\eta},\frac{Q^2}{\mu^2}\right) &\mapsto C_i\left(\frac{x}{\eta -i 0}, \frac{\xi}{\eta-i0},\frac{Q^2}{\mu^2}-i0\right)
= C_i\left(\frac{x}{\xi}(\omega+i0), \omega+i0,\frac{Q^2}{\mu^2}-i0\right).
   \end{align}
We have checked that the resulting one-loop CFs agree with the
explicit evaluation in Minkowski space in Ref.~\cite{Pire:2011st}.
%We will discuss analytic continuation in more detail in a later section.

The tree-level CFs are well-known since the pioneering works \cite{Ji:1996nm,Radyushkin:1996nd}
\begin{align}\label{treeCFs}
C_\perp^{(0)}(z,\omega ) &= \frac{\omega}{1 - z} - \frac{\omega}{1 + z}\,, \qquad  C_L^{(0)}(z,\omega) = 0\,.
\end{align}
We find it convenient to present the results for loop corrections in the following generic form
to emphasize the symmetries and the structure of singularities:
\begin{align}\label{loopCFs}
C_i^{(k)}\left(z,\omega, \frac{ Q^2}{\mu^2} \right) &=\frac{C_F\,\omega}{1-z} \Big[ A_i^{(k)}\left(z,\omega,\frac{ Q^2}{\mu^2}\right)
+A_i^{(k)}\left(z,-\omega,\frac{ Q^2}{\mu^2}\right)\Big]
\notag\\&\quad
+ \frac{C_F}{\omega-z} \Big[B_i^{(k)}\left(z,\omega,\frac{ Q^2}{\mu^2} \right) 
- B_i^{(k)}\left(-z,-\omega,\frac{ Q^2}{\mu^2} \right) 
\Big]
- (z\leftrightarrow -z)\,.
\end{align}
The one-loop results have been available for a long time \cite{Ji:1997nk,Mankiewicz:1997bk,Belitsky:2005qn,Pire:2011st}:
\begin{align}  \label{oneloopCFs}
A_\perp^{(1)}\left(z,\omega,\frac{Q^2}{\mu^2}\right) &= 
\ln^2(1\!-\!z)-\ln^2(1\!-\!\omega) -\frac32\ln(1\!-\!z) +3\ln(1\!-\!\omega) -\frac92
\notag\\&\quad
+ 2 \ln \frac{Q^2}{\mu^2} \Big[\ln(1\!-\!z) - \ln(1\!-\!\omega) + \frac34\Big], % \biggr\},
\notag\\
B_\perp^{(1)}\left(z,\omega,\frac{Q^2}{\mu^2}\right) &= 
- \frac{1+\omega}2\Big[ \ln^2(1\!-\!z)-\ln^2(1\!-\!\omega)\Big] 
 + 3\omega\Big[\ln(1\!-\!z)-\ln(1\!-\!\omega)\Big]
\notag\\&\quad
- \ln \frac{Q^2}{\mu^2} (1+\omega) \Big[\ln(1\!-\!z)- \ln(1\!-\!w)\Big],
\notag\\
A_L^{(1)}\left(z,\omega,\frac{Q^2}{\mu^2}\right) & =0\,,  
\notag\\
B_L^{(1)}\left(z,\omega,\frac{Q^2}{\mu^2}\right) &=  2\Big[\ln(1\!-\!z) - \ln(1\!-\!w)\Big] \,.
\end{align}
It is important that, despite the factor $1/(\omega -z)$ present in Eq.~\eqref{loopCFs}, 
the one-loop CFs are analytic functions at $z=\omega$. They are also analytic functions
in the limit $\lambda\to 0$, rescaling
$Q^2\mapsto \lambda Q^2$, $\omega\mapsto \omega/\lambda$, $z\mapsto z/\lambda$.
The latter property ensures that collinear factorization holds at the kinematic point $q_1^2 + q_2^2 = 0$, as
expected from the leading regions analysis~\cite{Belitsky:2005qn}.  
We will find that the two-loop CFs have the same analytic properties.

The two-loop CFs contain contributions of three different color structures. We choose them as follows:
\begin{align}
C_{i}^{(2)}\Big(z, \omega,\frac{Q^2}{\mu^2}\Big) & =
 C_F \biggl[\beta_0 C_{i}^{(2,\beta)}\Big(z, \omega,\frac{Q^2}{\mu^2}\Big) + C_F C_i^{(2,P)}\Big(z, \omega,\frac{Q^2}{\mu^2}\Big)
+\frac{1}{N_c} C_i^{(2,NP)}\Big(z, \omega,\frac{Q^2}{\mu^2}\Big)\biggr],
\label{color}
\end{align}
and use the same color decomposition for the $A_i^{(2)}$ and $B_i^{(2)}$ functions defined in Eq.~\eqref{loopCFs}, apart
from the overall $C_F$ factor.
The two-loop results turn out to be rather lengthy so that we write them 
separating the renormalization group (RG) logarithms, with the notation  
\begin{align}\label{RGlogs}
A_i^{(2)}\Big(z,\omega,\frac{Q^2}{\mu^2}\Big) &=
A_i^{(2)}\left(z,\omega\right) +\ln\Big(\frac{Q^2}{\mu^2}\Big) A_{i,\ln}^{(2)}\left(z,\omega\right)
+\ln^2\Big(\frac{Q^2}{\mu^2}\Big) A_{i,\ln^2}^{(2)}\left(z,\omega\right)\,,
\end{align}
and similarly for $B_i^{(2)}$.

%%%%%%%%%%%%%%%%%%%%%%%%%%%%%%%%%%%%%%%%%%%%%%%%%%%%%%%%%%%%%%%%%%%%%%%%%%%%%%%%%%%%%%%%%%%%%%%%%%%%%%%%%%%%%%%%%%%%%%

                \section{General framework }\label{sec:framework}

%%%%%%%%%%%%%%%%%%%%%%%%%%%%%%%%%%%%%%%%%%%%%%%%%%%%%%%%%%%%%%%%%%%%%%%%%%%%%%%%%%%%%%%%%%%%%%%%%%%%%%%%%%%%%%%%%%%%%%

One can consider, formally, the two-photon reactions in a generic $4-2\epsilon$-dimensional theory. All definitions in
Sec.~\ref{sec:general} can be taken over without modifications except for that the CFs acquire an $\epsilon$-dependence so that
     $$C_i(z,\omega, Q^2/\mu^2,a_s)\mapsto C_i(z,\omega, Q^2/\mu^2,a_s,\epsilon)\,.$$
Hence their perturbative expansion involves $\epsilon$-dependent coefficients:
\begin{align}
C(a_s,\epsilon)& =C^{(0)} + a_s\, C^{(1)}(\epsilon) +  a_s^2\, C^{(2)}(\epsilon)+ \mathcal{O}(a_s^3)\,,
\notag\\
C^{(k)}(\epsilon)& = C^{(k)} + \epsilon\, C^{(k,1)} + \epsilon^2 C^{(k,2)} + \mathcal{O}(\epsilon^3)\,.
\label{eq:generic-d}
\end{align}
Note that the tree-level CF $C^{(0)}$ does not depend on $\epsilon$.

We are interested in the CFs in four dimensions as a function of the coupling,
but as an intermediate step will
calculate $C_\ast = C(\alpha_s^*,\epsilon)$ on the line in the $(\epsilon, \alpha_s)$ plane
where $\beta(\alpha_s^\ast)=0$ so that $\alpha_s^\ast =\alpha_s^\ast(\epsilon)$ (Wilson-Fisher fixed point)
or, equivalently,
\begin{align}
\epsilon_\ast &= \epsilon_\ast(a_s) = -\big(\beta_0 a_s + \beta_1 a_s^2 + \ldots\big)\equiv -\bar\beta(a_s)\,, \qquad
\beta_0 = \frac{11}{3}N_c -\frac23 n_f\,,
\label{epsilon*}
\end{align}
with $N_c$ and $n_f$ being the numbers of colors and light flavors, respectively.
Trading the $\epsilon$-dependence for the $a_s^\ast$ dependence one can write the
CFs on this line as an expansion in the  coupling alone,
\begin{align}
C_\ast(a_s) & = C(a_s,\epsilon_\ast)= C^{(0)} +a_s C_\ast^{(1)} +a_s^2 C_\ast^{(2)}+\mathcal{O}(a_s^3)\,,
\label{eq:critical-expand}
\end{align}
where, obviously,
\begin{align}
   C_\ast^{(1)} &= C^{(1)}\,,
\qquad
   C_\ast^{(2)}  = C^{(2)} - \beta_0 C^{(1,1)}\,,
\label{eq:d=4-restored1}
\end{align}
or, equivalently
\begin{align}
 C^{(1)} & = C_\ast^{(1)}\,,
\qquad
 C^{(2)}  = C_\ast^{(2)} +\beta_0 C^{(1,1)}\,.
\label{eq:d=4-restored}
\end{align}
The rationale for organizing the calculation in this way is that QCD at the Wilson-Fischer critical point is
conformally invariant~\cite{Fortin:2012hn,Braun:2018mxm}. Conformal symmetry allows one to obtain the coefficients  $C_\ast^{(2)}$
from the known results from DIS avoiding explicit calculation. To restore the result in 4 dimensions 
one also needs to know terms of order $\epsilon$ in the one-loop CFs. This additional calculation is, however, rather simple.

It is straightforward to continue this construction to higher orders.
The general statement is that the $\ell$-loop
off-forward CFs in QCD in $d=4$ in the $\overline{\text{MS}}$ scheme can be obtained from the corresponding result in conformal
theory (alias from the corresponding CFs in the forward limit),
adding terms proportional to the QCD beta-function. Such extra terms require the calculation of the corresponding
$(\ell-1)$-loop off-forward CFs in $d=4-2\epsilon$ dimensions expanded to order $\epsilon^{\ell-1}$. 

%####################################################################################################################################

                            \subsection{Conformal OPE}\label{sect:COPE}

%####################################################################################################################################

The OPE for the product of currents has a generic form, schematically
\begin{align}
T\{j^{\rm em}_\mu(x)j^{\rm em}_\nu(0)\} &= \sum_{N,k} C_{N,k} \partial_+^k\mathcal O_N(0)\,,
\end{align}
where $\mathcal O_N(0)$ are local operators of increasing dimension and $C_{Nk}$ are the corresponding CFs.
The power of conformal symmetry is that it allows one to restore the contributions of all operators containing
total derivatives from the ones without total derivatives, i.e. restore $C_{N,k}$ from $C_{N,0}$ using
conformal algebra~\cite{Ferrara:1972uq}.

Retaining the contributions of twist-two vector operators only, the result reads~\cite{Braun:2020yib}
\begin{align}\label{COPE}
\text{T}\,\big\{j^\mu(x_1) j^\nu(x_2)\big\} & =
\sum_{N,\text{even}}\frac{\mu^{\gamma_N}}{(-x_{12}^{2}+i0)^{t_N}}
 \int_0^1 du
 \Biggl\{ -\frac12 A_N(u) \left( g^{\mu\nu} - \frac{2x_{12}^\mu x_{12}^\nu}{x_{12}^2}
 \right) + B_N(u) g^{\mu\nu}
 \notag\\
&\quad
 +C_N(u) x_{12}^\nu\partial_1^\mu - C_N(1-u) x_{12}^\mu
 \partial_2^\nu
% \notag\\
% &\quad
+D_N(u)x_{12}^2 \partial_1^\mu\partial_2^\nu \Biggr\}
\mathcal O_N^{x_{12}\ldots x_{12}}(x_{21}^u)\,,
\end{align}
where
\begin{align}
  \partial^\mu_k = \frac{\partial}{\partial x^\mu_k}, \qquad\quad x_{12} = x_1-x_2\,, %\qquad\quad \bar u = 1-u\,, 
  \qquad\quad x_{21}^u=(1-u) x_2+ u x_1\,,
\end{align}
and
\begin{align}
 \mathcal O_N^{x\ldots x}(y) = x_{\mu_1}\ldots x_{\mu_N}  \mathcal O_N^{\mu_1\ldots\mu_N}(y)\,.
\end{align}
Further, $\mathcal O_N^{\mu_1\ldots\mu_N}(y)$ are the leading-twist conformal operators
that transform in the proper way under conformal transformations
\begin{align}
   [\mathbb{K}_\mu, \mathcal O_N^{x\ldots x}(y)] =
   \left( 2y_\mu  y^\nu \frac{\partial}{\partial y^\nu}-y^2 \frac{\partial}{\partial y^\mu}
   +2\Delta_N y_\mu  +2y^\nu\left(x_\mu \frac{\partial}{\partial x^\nu} -
   x_\nu \frac{\partial}{\partial x^\mu}\right)\right) \mathcal O_N^{x\ldots x}(y).
\end{align}
Here and below, $N$ is the spin and $\Delta_N$ the scaling dimension of $\mathcal O_N^{\mu_1\ldots\mu_N}$,
$\Delta_N=d_\ast+N-2+\gamma_N$ where $d_\ast = 4-2\epsilon_\ast$,
$\gamma_N=\gamma_N (a_s)$ is the anomalous dimension,
$t_N= 2 -\epsilon_\ast -\frac12\gamma_N(a_s)$ is the twist and
$j_N = N + 1 -\epsilon_\ast +\frac12\gamma_N(a_s)$ is the conformal spin.
We have separated in Eq.~\eqref{COPE} the scale factor $\mu^{\gamma_N}$ to make the invariant functions
$A_N(u),\ldots, D_N(u)$  dimensionless.
Note that only vector operators with even spin $N$ contribute to the expansion.

Conformal invariance and current conservation $\partial^\mu j_\mu=0$ constrain  the functional form
of the invariant functions $A_N(u),\ldots, D_N(u)$ in Eq.~\eqref{COPE}
and also lead to certain relations between them. % the invariant functions $A_N(u),\ldots, D_N(u)$ in Eq.~\eqref{COPE}.
One obtains~\cite{Braun:2020yib}
\begin{align}
A_N(u) = a_N(a_s)\, u^{j_N-1}(1-u)^{j_N-1}\, , && B_N(u) = b_N(a_s) u^{j_N-1} (1-u)^{j_N-1}\, .
\label{AB}
\end{align}
Explicit expressions for $C_N(u)$ and $D_N(u)$ can be found in Ref.~\cite[Eq.~(3.12)]{Braun:2020yib}
and do not involve new parameters. Thus the OPE of the product of two conserved spin-one currents in a
generic conformal theory involves two constants, $a_N(a_s)$ and $b_N(a_s)$, for each (even) spin~$N$.
In QCD the expansion  of  $a_N(a_s)$ starts at order $\mathcal{O}(a_s)$.

For the matrix elements between states with the same momentum (forward scattering), the
position of the operators  on the r.h.s of Eq.~\eqref{COPE} is irrelevant and the integration
over the $u$-variable can be taken explicitly.
It is convenient to fix the normalization of the operators such that
\begin{align}
   \mathcal O_N^{\mu_1\ldots\mu_N}(0) = i^{N-1}\bar q(0) \gamma^{\{\mu_1} D^{\mu_2}\ldots D^{\mu_N\}} q(0)
   +\,\text{total~derivatives}\,,
\label{eq:normalization}
\end{align}
where $D^{\mu} = \partial^{\mu} + i g A^{\mu}$ and
$\{\ldots\}$ denotes the symmetrization of all enclosed Lorentz indices and the subtraction
of traces. In this way the forward matrix elements of these operators can be identified with the moments of
quark parton distributions (PDFs)
\begin{align}
   \langle p| \mathcal O_N^{\mu_1\ldots\mu_N}(0)|p\rangle = p^{\{\mu_1}\ldots p^{\mu_N\}} f_N \,.
\end{align}
With this normalization one obtains~\cite{Braun:2020yib}
\begin{flalign}
T_{\mu\nu}(p,q) &\equiv i\int d^dx\, e^{-iqx}\langle p|T(j_\mu(x)j_\nu(0)|p\rangle
\notag\\
&=
\sum_{N,\text{even}}\!\!
\frac{f_N}{x_B^N}
%\!\left(\frac{2pq}{Q^2}\right)^N\!\!\!
\left(\frac{\mu}{Q}\right)^{\gamma_N}
\!\biggl[
\left(-g_{\mu\nu} + \frac{q_\mu q_\nu}{q^2}\right)c_{1N}(a_s)
+\frac{(q_\mu\!+\!2x_Bp_\mu)(q_\nu\!+\!2x_Bp_\nu)}{Q^2} c_{2N}(a_s)
\biggr],
\end{flalign}
where $x_B = Q^2/(2qp)$ is the Bjorken scaling variable and
\begin{align}
 c_{1N} &= i^N\pi^{d/2} 2^{\gamma_N} B(j_N,j_N) \frac{\Gamma(N+\gamma_N/2)}{\Gamma(t_N)}
\left( \frac{t_N-1}{2t_N}a_N-b_N \right),
\notag\\
c_{2N} & = i^N\pi^{d/2} 2^{\gamma_N} B(j_N,j_N) \frac{\Gamma(N+\gamma_N/2)}{\Gamma(t_N)}
\left(-b_N + \frac{2N+d -t_N-1}{2t_N} a_N\right).
\end{align}
Here and below  $B(j_N,j_N)$ is the Euler Beta function.

Comparing this expression with the usual expansion for the DIS structure functions, see e.g.~\cite{Vermaseren:2005qc},
we can identify
\begin{align}
c_{2N}(a_s)\left(\frac{\mu}{Q}\right)^{\gamma_N}  & = C_{2}^\DIS\left(N,\frac{Q^2}{\mu^2},a_s,\epsilon_\ast\right),
\notag\\
c_{1N}(a_s)\left(\frac{\mu}{Q}\right)^{\gamma_N}  & = C_{2}^\DIS\left(N,\frac{Q^2}{\mu^2}, a_s,\epsilon_\ast\right)-
C_{L}^\DIS\left(N,\frac{Q^2}{\mu^2},a_s,\epsilon_\ast\right)
\equiv C_{1}^\DIS\left(N,\frac{Q^2}{\mu^2},a_s,\epsilon_\ast\right).
\end{align}
where $C_2^\DIS$ and $C_L^\DIS$ are the familiar CFs for the structure functions $F_2$ and $F_L$, respectively, (in $4-2\epsilon_\ast$ dimensions)
that are known to third order in the QCD coupling. With this identification,
the structure of the OPE for the product of two vector currents in conformal QCD is completely fixed.

For the off-forward case there are two modifications.
First, the position of the operator $\mathcal O_N(ux)$ in Eq.~\eqref{COPE} becomes relevant since
\begin{align}
 \langle p_2|\mathcal O_N(x_{21}^u )|p_1\rangle= e^{ i (x_{21}^u\cdot\Delta)}\langle p_2|\mathcal O_N(0)|p_1\rangle\,,
\end{align}
producing a $u$-dependent shift of the momentum in the Fourier integral.
Second, the matrix element becomes
more complicated. It can be parameterized as
\begin{align}
\langle p_2|n^{\mu_1} \ldots n^{\mu_N} \mathcal O_{\mu_1\ldots\mu_N}(0)|p_1\rangle & =
\sum_k \left(-\frac12\right)^k f_N^{(k)} p_+^{N-k} \Delta_+^k =  p_+^N f_N(\xi)\,,
\notag\\
f_N(\xi) &\equiv \sum_k f_N^{(k)}\xi^k\,, \qquad f_N^{(0)}= f_N^{\rm DIS}\,.
\label{fN(xi)}
\end{align}
The leading-twist CFFs $\mathcal{F}_1(\xi,\eta, \Delta^2,Q^2)$ and $\mathcal{F}_L(\xi,\eta, \Delta^2,Q^2)$~\eqref{F12L} can be separated
by taking  $g_\perp^{\mu\nu}\mathcal{A}_{\mu\nu}$ and $\Delta^\mu\mathcal{A}_{\mu\nu}\Delta^\nu$ projections, respectively.
In this way the (complicated) $C_N(u)$ and $D_N(u)$ terms in \eqref{COPE} drop out, and one obtains after a short calculation
\begin{align}
  \mathcal{F}^\ast_\perp(\xi,\eta,Q^2) &=
\sum_N f_N(\xi) \eta^{-N}
\frac{C_1^\DIS(N,\tfrac{Q^2}{\mu^2}, a_s,\epsilon_\ast) }{(1+w)^{\tfrac12\gamma_N+N}}
		{}_2F_1\left(\tfrac12\gamma_N+N, j_N, 2j_N, \frac{2w}{1+w}\right),
\label{DDVCS-T}
\\
  \mathcal{F}^\ast_L(\xi,\eta,Q^2) &=
%- \sqrt{1-\omega^2}%\frac{1}{\omega^2 q^2 }
\sum_N f_N(\xi) \eta^{-N}
\frac{C_L^\DIS(N,\tfrac{Q^2}{\mu^2}, a_s,\epsilon_\ast)}{(1+w)^{\tfrac12\gamma_N+N+1}}
		{}_2F_1\left(\tfrac12\gamma_N+N+1, j_N, 2j_N, \frac{2w}{1+w}\right),
\label{DDVCS-L}
\end{align}
where the superscript ${}^\ast$ indicates that these results refer to QCD at the critical point. Hereafter we do not show the dependence of the
CFFs on $\Delta^2$, which only enters through the matrix elements and does not affect CFs.

%################################################################################################################################

                  \subsection{Coefficient functions in momentum fraction space: master equation}

%#################################################################################################################################

As the next step, we have to find a way to obtain the CFs in momentum fraction space $C_{1,L}(z,w)$  starting from these
expressions.
To this end one needs to write the  GPD in terms of the matrix elements of conformal operators, which is difficult. 
The form of these operators is determined by the generator of special conformal transformation, which is modified 
in an interacting theory compared to the ``canonical'' expression, and is rather complicated in QCD in 
the $\overline{\rm MS}$ scheme~\cite{Braun:2016qlg}. 
The way out~\cite{Braun:2017cih,Braun:2020yib} is to go over to a different, ``rotated'' renormalization scheme at the intermediate step,
\begin{align}
 \mathcal{F}_i(\xi,\eta) = \int_{-1}^1\!\frac{dx}{\xi} \, C_i\left(\frac{x}{\eta},\omega, \frac{Q^2}{\mu^2}\right) F_q(x,\xi,\mu^2)
  = \int_{-1}^1\!\frac{dx}{\xi} \, \mathbf C_i\left(\frac{x}{\eta},\omega, \frac{Q^2}{\mu^2}\right) \mathbf F_q(x,\xi,\mu^2)\,,
\label{rot1}
\end{align}
where
\begin{align}
\mathbf{F}_q(x,\xi)  & =  \int_{-1}^1 \frac{dx'}{\xi} \mathrm{U}(x,x',\xi) F_q(x',\xi)\,,
\notag\\[2mm]
C_i\left(\frac{x}{\eta},\omega, \frac{Q^2}{\mu^2}\right)
& = \int_{-1}^1 \frac{dx'}{\xi}  \mathbf C_i \left(\frac{x'}{\eta},\omega, \frac{Q^2}{\mu^2}\right) \mathrm{U}(x',x,\xi)\,.
\label{F-T-C}
\end{align}
The operator
\begin{align}
   \mathrm{U} = e^{\mathbb{X}}\,, \qquad \mathbb{X}(a_s)  =  a_s \mathbb{X}^{(1)}+ a_s^2 \mathbb{X}^{(2)}+\ldots\,,
\label{similarity2}
\end{align}
is defined in such a way  that the ``rotated'' generators of conformal transformations $\mathbf S_{\pm,0} = \mathrm{U}{S}_{\pm,0} \mathrm{U}^{-1}$
are given entirely in terms of the ``rotated'' evolution kernel \eqref{RGE} $\mathbf{H} = \mathrm U \, \mathbb{H}\, \mathrm U^{-1}$.
In the light-ray operator (position space) representation
\begin{subequations}
\label{Sbold}
\begin{align}\label{translationbold}
   \mathbf{S}_- &= {S}_-^{(0)}\,,
\\
\label{dilatationbold}
   \mathbf{S}_0\, &= {S}_0^{(0)} -\epsilon_\ast + \frac12\mathbf{H}\,,
\\
   \mathbf{S}_+   &= {S}_+^{(0)} + (z_1+z_2)\left(-\epsilon_\ast+ \frac12  \mathbf{H}\right)\,,
\label{special-conformalbold}
\end{align}
\end{subequations}
where
\begin{align}
 {S}_-^{(0)} = -\partial_{z_1} -\partial_{z_2}\,, &&
{S}_0^{(0)} = z_1\partial_{z_1} + z_2\partial_{z_2}+2\,, &&
 {S}_+^{(0)} =  z_1^2\partial_{z_1} + z_2^2\partial_{z_2}+2(z_1+z_2)\,,
\label{Scanonical}
\end{align}
are the canonical generators.
Explicit expressions for $\mathbb{X}^{(1)}$ and $\mathbb{X}^{(2)}$ in the position-space representation can be found in~\cite{Braun:2017cih}.

One obtains \cite[Eq.(3.49)]{Braun:2020yib} for the GPD at $\xi=1$ in the ``rotated'' scheme
\begin{align}
\label{Frotated2}
\mathbf{F}(x,\xi =1)=\frac1{4}\sum_{N} \frac{\sigma_N\, \omega_N}{2^{N-1} (N-1)!} \,f_{N}(\xi=1) P^{(\lambda_N)}_{N-1}(x)
\,,
\end{align}
where
\begin{align}
P^{(\lambda_N)}_{N-1}(x)=\left(\frac{1-x^2}4\right)^{\lambda_N-\frac12} C_{N-1}^{\lambda_N}(x)\,, &&
 \lambda_N = \frac32 -\epsilon_\ast + \frac12 \gamma_N(a_s)\,,
\label{lambdaN}
\end{align}
$C^{\lambda}_{N}$ are Gegenbauer polynomials,  $\sigma_N(a_s)$ are eigenvalues
of the rotation operator $\mathrm U$
\begin{align}
  \mathrm U\, z_{12}^{N-1} = \sigma_N  z_{12}^{N-1}\,,  && \sigma_N(a_s) = 1 + a_s\sigma_N^{(1)} +  a^2_s\sigma_N^{(2)}+ \ldots
\end{align}
and
\begin{align}
 \omega_N & =
  \frac{(N-1)!\, \Gamma(2j_N)\Gamma(2\lambda_N)}{\Gamma (\lambda_N+\frac12)\Gamma(j_N)\Gamma(N-1+2\lambda_N)}\,.
\label{omegaN}
\end{align}
The restriction to $\xi=1$ is due to the well-known problem caused by non-uniform convergence of a sum
representation for GPDs in the DGLAP region $\xi < |x|$. This result is sufficient, however, because the
CFs only depend on the ratios of scaling variables $x/\eta, \xi/\eta$ so that for our purposes we can set  $\xi=1$ and eliminate
the DGLAP region completely. Using this expression in Eq.~\eqref{rot1} and comparing the result with the
expansion in \eqref{DDVCS-T}, \eqref{DDVCS-L} one obtains
\begin{align}
\int_{-1}^1 \!\!dx\,\mathbf C_\perp\left(\omega x,\omega, \tfrac{Q^2}{\mu^2}\right) P^{(\lambda_N)}_{N-1}(x)
&=
\frac{C_1^\DIS(N,\tfrac{Q^2}{\mu^2}, a_s,\epsilon_\ast)}{(1\!+\!\omega)^{\tfrac12\gamma_N}} \Big(\frac{2\omega}{1\!+\!w}\Big)^N
\!\!{}_2F_1\left(\tfrac12\gamma_N+N, j_N, 2j_N, \frac{2\omega}{1\!+\!\omega}\right)
\notag\\&\quad
\times
  \frac{2 \Gamma (\lambda_N+\frac12)\Gamma(j_N)\Gamma(N-1+2\lambda_N)}{\sigma_N  \Gamma(2j_N)\Gamma(2\lambda_N)},
\notag\\
\int_{-1}^1\!\! dx\,\mathbf C_L\left(\omega x,\omega, \tfrac{Q^2}{\mu^2}\right) P^{(\lambda_N)}_{N-1}(x)
&=
\frac{C_L^\DIS(N,\tfrac{Q^2}{\mu^2}, a_s,\epsilon_\ast)}{(1\!+\!\omega)^{1+ \tfrac12\gamma_N}}
\!\!\Big(\frac{2\omega}{1\!+\!w}\Big)^N \!\! {}_2F_1\left(\tfrac12\gamma_N\!\!+\!\!N\!\!+\!\!1, j_N, 2j_N, \frac{2\omega}{1\!+\!\omega}\right)
\notag\\&\quad
\times
  \frac{2 \Gamma (\lambda_N\!+\!\frac12)\Gamma(j_N)\Gamma(N\!-\!1\!+\!2\lambda_N)}{\sigma_N  \Gamma(2j_N)\Gamma(2\lambda_N)}.
\label{mastereq}
\end{align}
DVCS corresponds to $\omega=1$ in which case
\begin{align}
 \frac{1}{(1+\omega)^{\tfrac12\gamma_N+N}} {}_2F_1\left(\tfrac12\gamma_N+N, j_N, 2j_N, \frac{2\omega}{1+\omega}\right) \mapsto
\frac{1}{2^{\tfrac12\gamma_N+N}}\frac{\Gamma(\dhalf-1)\Gamma(2j_N)}{\Gamma(j_N+\dhalf-1) \Gamma(j_N)}
\end{align}
and the first equation in \eqref{mastereq} reduces to the corresponding expression in Ref.~\cite{Braun:2020yib} apart from the
$2^{-\gamma_N/2}$ factor, which is due to a different choice of the hard scale: in~\cite{Braun:2020yib} $Q^2 = -q_1^2$.

It remains to solve the equations~\eqref{mastereq} to obtain the CFs in ``rotated'' scheme, and apply a finite renormalization
\eqref{F-T-C} to arrive at the final expressions in the $\overline{\rm MS}$  scheme.
In the next section we outline the general procedure for this calculation.

%################################################################################################################################

                       \subsection{Solution ansatz and invariant kernels}
                       \label{sec:solansatzinvkernel}

%#################################################################################################################################

The scale-dependent terms $\sim \ln \tfrac{Q^2}{\mu^2}$ in the CFs can be restored from the renormalization group equations, see App.~\ref{App:ScaleDependence}.
This task is easy, so that  we concentrate on the case $\mu^2=Q^2$. Hereafter 
$\mathbf{C}_i\left(\omega x,\omega\right) \equiv \mathbf{C}_i\left(\omega x,\omega, 1\right)$.

At leading order $\lambda_N=3/2$ and the functions $P_{N-1}^{(\lambda_N)}\left(x\right)$ form an orthonormal system. Hence 
one can write the CFs, at least formally, as a series over these functions. 
Beyond the leading order this cannot be done, because  
$P_{N-1}^{(\lambda_N)}\left(x\right)$ with different $N$ are not orthogonal with any simple weight function. 
However, these functions are eigenfunctions of the Casimir operator corresponding to the
``rotated'' conformal generators~\eqref{Sbold}, and, therefore, also eigenfunctions of the (exact) ``rotated'' evolution kernel
\begin{align}\label{Hxxprime}
   \int dx' \, \mathbf{H}(x,x') \, P^{(\lambda_N)}_{N-1}(x') = \gamma_N P^{(\lambda_N)}_{N-1}(x)\,.
\end{align}
This property suggests the following ansatz for the CFs:
\begin{align}\label{CK-ansatz}
\mathbf{C}_i\left(\omega x,\omega\right)=  \int_{-1}^1 dx'\, c_i(\omega,x') K_i\left(x',x,\omega\right)\,,
\end{align}
where $c_i(\omega,x)$ are certain weight functions (see below), and $K_i(x,x',\omega)$ are $\mathrm{SL}(2)$-invariant operators, $[K_i,\mathbf S_{\pm,0}]=0$.
Since the polynomials $P_{N-1}^{(\lambda_N)}\left(x\right)$ are eigenfunctions of the quadratic Casimir operator, they are also
eigenfunctions of {\it any} $\mathrm{SL}(2)$-invariant operator, i.e.
\begin{align}\label{KPN}
\int dx'\,K_i\left(x',x,\omega\right)\,P_{N-1}^{(\lambda_N)}(x') = K_i\left(N,\omega\right)\,P_{N-1}^{(\lambda_N)}(x)\,.
\end{align}
Using the above ansatz \eqref{CK-ansatz} one obtains for the integrals on the l.h.s. of Eqs.~\eqref{mastereq}
\begin{align}\label{CPN}
\int_{-1}^1 dx \,  \mathbf{C}_i(\omega x,\omega)\, P_{N-1}^{(\lambda_N)}(x)
&=
\int_{-1}^1 dx  \int_{-1}^1 dx'\,c_i(\omega,x') K_i(x',x,\omega)\, P_{N-1}^{(\lambda_N)}(x)
\notag\\&=
K_i(N,\omega)\,\int_{-1}^1 dx \,c_i(\omega,x ) P_{N-1}^{(\lambda_N)}(x)\,.
\end{align}
The weight functions $c_i(\omega,x)$ can be fixed by the requirement that they lead to sufficiently simple
resulting expressions for the eigenvalues  $K_i(N,\omega)$ of the invariant kernels.
We require that (cf.~\eqref{CPN})
\begin{align}
\int_{-1}^1 dx\,  c_\perp(\omega,x) P_{N-1}^{(\lambda_N)}(x) &=
\int_{-1}^1 dx\, \biggl\{\frac{\omega}{(1 - \omega x)^{1+\frac12\gamma_N}} - \frac{\omega}{ (1+\omega x)^{1+\frac12 \gamma_N}}
\biggr\}P_{N-1}^{(\lambda_N)}(x),
\notag\\
\int_{-1}^1 dx\, c_L(\omega,x) P_{N-1}^{(\lambda_N)}(x)&=
\int_{-1}^1 dx\, \biggl\{\frac{\omega}{(1 - \omega x)^{2+\frac12\gamma_N}} - \frac{\omega}{ (1+\omega x)^{2+\frac12\gamma_N}}\biggr\}
P_{N-1}^{(\lambda_N)}(x).
\label{weights}
\end{align}
The functions $c_i(x,\omega)$ themselves can easily be restored from these expressions by observing that the anomalous
dimension $\gamma_N$ is, by definition, the eigenvalue of the (exact) ``rotated'' evolution kernel \eqref{Hxxprime}, so that effectively
\begin{align}
{(1\pm \omega x)^{-n-\frac12\gamma_N}}\mapsto
\sum_{k=0}^\infty\frac{1}{k!} \int_{-1}^1 dx'\frac{\ln^k(1 \pm \omega x')}{(1\pm  \omega x')^n} \left(-\frac12 \mathbf H(x',x)\right)^k\,,
\label{weights2}
\end{align}
The integrals on the r.h.s. of \eqref{weights} can be taken explicitly,
\begin{align}
\int_{-1}^1 dx\, c_\perp(\omega,x) P_{N-1}^{(\lambda_N)} (x) & =
\frac{1\!+\!(-1)^N}{2} \Big( \frac{2\omega}{1\! +\! \omega}\Big)^N \Big( \frac{1}{1 \!+\! \omega} \Big)^{\tfrac12 \gamma_N}
\frac{2 \Gamma(\lambda_N + \tfrac12 ) \Gamma(N - 1 + 2\lambda_N) \Gamma(j_N)}{\Gamma(2\lambda_N) \Gamma(2 j_N)}
\nonumber \\& \quad
\times \frac{\Gamma(N + \tfrac12 \gamma_N)}{\Gamma(N) \Gamma(1 + \tfrac12 \gamma_N) }
	 {}_2F_1 \big(N + \tfrac12 \gamma_N, j_N , 2 j_N , \tfrac{2\omega}{1 + \omega} \big) ,
\notag\\
\int_{-1}^1 dx\, c_L(\omega,x) P_{N-1}^{(\lambda_N)} (x) & =
(1\!+\!(-1)^N) \Big( \frac{2\omega}{1 \!+\! \omega}\Big)^N\!\! \Big( \frac{1}{1\! +\! \omega} \Big)^{1+\tfrac12 \gamma_N}
\frac{\Gamma(\lambda_N + \tfrac12 ) \Gamma(N \!-\! 1\! +\! 2\lambda_N) \Gamma(j_N)}{\Gamma(2\lambda_N) \Gamma(2 j_N)}
\nonumber \\& \quad
\times \frac{\Gamma(N + 1+\tfrac12 \gamma_N)}{\Gamma(N) \Gamma(2 + \tfrac12 \gamma_N) }
	 {}_2F_1 \big(N + 1+ \tfrac12 \gamma_N, j_N , 2 j_N , \tfrac{2\omega}{1 + \omega} \big) .
\end{align}
Comparing these expressions with \eqref{mastereq} one obtains
\begin{align}
K_\perp(\omega, N)  &=
  \frac{\Gamma(N) \Gamma(1 + \tfrac12 \gamma_N) }{\sigma_N \Gamma(N + \tfrac12 \gamma_N)} C_1(N,\tfrac{Q^2}{\mu^2}, a_s,\epsilon_\ast)
  	\equiv K_\perp(N)\,,
\notag\\
K_L(\omega, N)  &=
  \frac{\Gamma(N) \Gamma(2 + \tfrac12 \gamma_N) }{\sigma_N \Gamma(N + 1+ \tfrac12 \gamma_N)} C_L^\DIS(N,\tfrac{Q^2}{\mu^2}, a_s,\epsilon_\ast)
  	\equiv K_L(N)\,.
\label{spectrum-KN}
\end{align}
Remarkably, with the choice \eqref{weights}, the invariant kernels $K_{\perp,L}$ do not depend on $\omega$:
their spectrum is given directly in terms of moments of the DIS CFs and the
eigenvalues $\sigma_N$ of the rotation operator U.

Expanding all entries
in \eqref{weights2}, \eqref{spectrum-KN} in powers of the coupling constant
\begin{align}
  K_i(N) &= 1 + a_s  K_i^{(1)}(N) +  a_s^2  K_i^{(2)}(N) + \ldots,
&&
\mathbf H =  a_s \mathbf H^{(1)} + a_s^2  \mathbf H^{(2)} + \ldots
\end{align}
one obtains to one loop accuracy
\begin{align}
\gamma_N^{(1)} &= 4 C_F\biggl\{ 2 S_1(N) - \frac{1}{N(N+1)} -\frac32\biggr\},
\notag\\
  K_\perp^{(1)}(N) &= 2 C_F \biggl\{ 3  S_1(N) + \frac{1}{N(N+1)} - \frac92\biggr\},
\notag\\
  K_L^{(1)}(N) &= 4 C_F \frac{1}{N(N+1)}\,.
\label{K-1loop}
\end{align}

An $SL(2)$-invariant operator, i.e., an operator that commutes with the generators $\mathbf S_{\pm,0}$ of $SL(2,\mathbb{R})$ transformations,
is fixed uniquely by its spectrum. Therefore, Eq.~\eqref{K-1loop}
unambiguously  defines the operators $K_i^{(1)}$, $\mathbf H^{(1)}$, and, by virtue of Eq.~\eqref{CK-ansatz}, also the CFs
$\mathbf{C}_i^{(1)}(x)$. Let
\begin{align}
[\mathcal H_+ f] (z_1,z_2) & =\int_0^1d\alpha \int^{\bar\alpha}_0 d\beta \, f(z_{12}^\alpha,z_{21}^\beta)\,,
\notag\\
[\widehat {\mathcal H} f](z_1,z_2) &=
\int_0^1\frac{d\alpha}{\alpha}\Big[2f(z_1,z_2)-\bar\alpha f(z_{12}^\alpha,z_2)-\bar\alpha f(z_1,z_{21}^\alpha)\Big]\,.
\label{eq:position-kernels}
\end{align}
These operators commute with the canonical generators $S^{(0)}_{\pm,0}$. Using $f(z_1,z_2) = z_{12}^{N-1}$ it is easy to check that
\begin{align}
  \mathcal H_+\, z_{12}^{N-1} =  \frac{1}{N(N+1)}  z_{12}^{N-1}\,,
&&
 \widehat {\mathcal H}\, z_{12}^{N-1} = 2 S_1(N)  z_{12}^{N-1}\,,
\end{align}
Thus
\begin{align}
\mathbf H^{(1)} &= 4 C_F\left(  \widehat {\mathcal H} - \mathcal H_+ -\frac32 \II \right)\,,
\notag\\
  K_\perp^{(1)}    &= 2 C_F \left( \frac32 \widehat {\mathcal H}  +  \mathcal H_+ - \frac92 \II \right),
\notag\\
  K_L^{(1)} &= 4 C_F  \mathcal H_+\,.
\label{K-1loopA}
\end{align}
The complete list of invariant kernels appearing in the two-loop calculation is given in Appendix~\ref{App:KernelList}.
They are sufficiently simple in the position-space representation and
can be transformed to the momentum fraction space, if desired.

The general procedure for the transformation from position to momentum space is as follows.
Let $R$ be an integral operator in position space%
\footnote{
Operators of this form commute with translations, $R T_a = T_a R$, where $[T_a f](z_1,z_2) = f(z_1+a,z_2+a)$.
As a consequence, $\xi$, which is the Fourier-conjugate variable to $z_1+z_2$, is conserved.}
\begin{align}
[Rf](z_1,z_2)=\int_0^1 d\alpha\int_0^{\bar\alpha} d\beta \, r(\alpha,\beta) f(z_{12}^\alpha,z_{21}^\beta)\,.
\label{R}
\end{align}
Going over to the  momentum fraction space $(z_1,z_2) \mapsto (x,\xi)$ corresponds to a Fourier transformation
in two variables  (cf. \eqref{H-def}),
%The corresponding expression in momentum fraction space $(z_1,z_2) \mapsto (x,\xi)$ (cf. \eqref{H-def}),
\begin{align}
f(z_1,z_2)&=\int dx \int d\xi\, e^{-i(\xi-x)z_1 -i(\xi+x)z_2} f_\xi(x)\,.
\end{align}
One obtains
\begin{align}
 [R_\xi f_\xi](x) &= \int_{-\infty}^\infty dx'\, r_\xi(x,x') \,f_\xi(x')\,,
\end{align}
where
\begin{align}
r_\xi(x,x')&=\int_0^1 d\alpha\int_0^{\bar\alpha} d\beta\, \delta\left(x'-(\alpha-\beta)\xi-(1-\alpha-\beta) x\right) \, r(\alpha,\beta)\,.
\end{align}

The expressions for the momentum fraction kernels $r_\xi(x,x')$ are in general much more involved as compared to their position-space
counterparts $r(\alpha,\beta)$. Fortunately, these expressions are not needed since the convolution integrals of the kernels with
the weight functions $c_i(x,\omega)$ \eqref{weights}, \eqref{weights2} can be calculated starting from the position space expressions directly:
\begin{align}\label{px}
\int dx' \frac{\ln^k(1-\omega x')}{(1-\omega x')^n} r_{\xi=1}(x',x)= \int_0^1 d\alpha\int_0^{\bar\alpha} d\beta \,r(\alpha,\beta)
\frac{\ln^k((1-x) \omega_{+-}^\alpha + (1+x) \omega_{-+}^\beta)}{\left((1-x) \omega_{+-}^\alpha + (1+x) \omega_{-+}^\beta\right)^n}\,,
\end{align}
where $\omega_+=(1+\omega)/2$, $ \omega_-=(1-\omega)/2$.
The $\alpha,\beta$ integral in the r.h.s. in \eqref{px} can be calculated with the help of
HyperInt package~\cite{Panzer:2014caa} for sufficiently large class of functions $r(\alpha,\beta)$. Note also that for $-1<\omega,x <1$ the
combination  $(1-x) w_{+-}^\alpha + (1+x) w_{-+}^\beta$ is positive in the whole integration domain.%
\footnote{If $R$ is given by a product of several operators of the form \eqref{R},  the right hand side of \eqref{px}
can be written as a multifold integral of the same type, cf.~\cite{Braun:2020yib}.}

Let us introduce a short-hand notation for the functions
\begin{align}
\Y_n^{(k)}(x,w) &= \frac{\omega (-1)^k}{2^k k!}\left(\frac{ \ln^k(1 - \omega x)}{(1 - \omega x)^n} -\frac{ \ln^k(1 + \omega x)}{(1 + \omega x)^n}\right)
\label{Lnk}
\end{align}
and for  the convolution
\begin{align}
(f\otimes R)(x) \equiv \int dx' f(x') R(x',x)\,.
\end{align}
In this notation, the one-loop CFs in the $\overline{\text{MS}}$ scheme take the form
\begin{align}\label{C1Lexplicit}
C^{(1)}_\perp(\omega x,\omega) &= \Y_1^{(0)} \otimes \Big( K_\perp^{(1)} +\mathbb X^{(1)} \Big) + \Y_1^{(1)} \otimes \mathbf H^{(1)}\,,
\notag\\
C^{(1)}_L(\omega x,\omega) &= \Y_2^{(0)} \otimes  K_L^{(1)}\,,
\end{align}
where $\mathbb X^{(1)}$ \eqref{similarity2} is given by~\cite{Braun:2017cih}:
\begin{align}
[\mathbb X^{(1)} f](z_1,z_2) & = 2 C_F \int_0^1 d\alpha\frac{\ln\alpha}{\alpha}\Big(2 f(z_1,z_2) - f(z_{12}^\alpha,z_2) -f(z_1,z_{21}^\alpha)\Big)\,.
\end{align}
Calculating the  convolution integrals in Eq.~\eqref{C1Lexplicit} we reproduce the known one-loop expressions
\cite{Ji:1997nk,Mankiewicz:1997bk,Belitsky:2005qn,Pire:2011st} collected in Eqs.~\eqref{oneloopCFs}.

%%%%%%%%%%%%%%%%%%%%%%%%%%%%%%%%%%%%%%%%%%%%%%%%%%%%%%%%%%%%%%%%%%%%%%%%%%%%%%%%%%%%%%%%%%%%%%%%%%%%%%%%%%%%%%%%%%%%%%%%%%%%%%%%%%%%%%%

                   \section{Two-loop coefficient functions}          \label{sec:twoloopCFs}

%%%%%%%%%%%%%%%%%%%%%%%%%%%%%%%%%%%%%%%%%%%%%%%%%%%%%%%%%%%%%%%%%%%%%%%%%%%%%%%%%%%%%%%%%%%%%%%%%%%%%%%%%%%%%%%%%%%%%%%%%%%%%%%%%%%%%%%

Beyond one-loop, reconstruction of the operator $K_i$ ($i=1,L$) from its eigenvalues $K_i(N)$ is more complicated
since, by definition, it commutes with deformed generators $\mathbf S_{\pm,0}$ \eqref{Sbold} 
which differ from the canonical generators~\eqref{Scanonical}.
It has been shown recently~\cite{Ji:2023eni} that any invariant operator $[\mathbf S_{\pm,0}^{(0)},K]=0$ can be cast in the form
\begin{align}\label{KTK}
K = T^{-1} \widehat K T ,  %&& T=\sum_{k=0}^\infty \frac1{k!} {\ln |z_{12}|^k} \left(\bar \beta(a_s) +\frac12\mathbf H(a_s)\right)^k\,.
\end{align}
where $\widehat K$ is the canonically invariant operator, $[S_{\pm,0}^{(0)},\widehat K]=0$, and the operator $T$ intertwines the 
deformed symmetry
generators $\mathbf S_{\pm,0}$, Eq.~\eqref{Sbold}, and the canonical ones, $S_{\pm,0}^{(0)}$, Eq.~\eqref{Scanonical},
\begin{align}
T\mathbf S_{\pm,0}= S_{\pm,0}^{(0)} T.
\end{align}
One finds~\cite{Ji:2023eni}
\begin{align}
T=\sum_{k=0}^\infty \frac{1}{k!} {\ln^k |z_{12}|} \left(\bar\beta(a_s) +\frac12\mathbf H \right)^k\,,
\end{align}
where $\bar \beta(a_s) = - \beta(a)/(2a)$ \eqref{epsilon*} is the usual QCD beta-function (in 4 dimensions) and $\mathbf H(a_s)$ is the (rotated) evolution kernel.

The eigenvalues of $K$ and $\widehat K$ are related by the so-called reciprocity relation~\cite{Dokshitzer:2005bf,Basso:2006nk}
\begin{align}
K(N) & =\widehat K \left (N + \bar\beta(a_s) + \frac12 \gamma_N(a_s) \right).
\label{reciprocity}
\end{align}
It turns out that the eigenvalues of $\widehat K(N)$ are invariant under the replacement $N\to -N-1$ at large $N$.
As a consequence, only special combinations
of the harmonic sums~\cite{Vermaseren:1998uu} --- the so called parity-invariant harmonic
sums~\cite{Dokshitzer:2006nm,Beccaria:2009vt} --- appear in the expansion of $\widehat K(N)$.

Expanding everything in powers of the coupling,
$K_i=a_s K_i^{(1)} +a_s^2 K_i^{(2)} +\cdots $, etc., one obtains
\begin{align}
K_i^{(1)} & = \widehat K_i^{(1)},
\notag\\
K_i^{(2)} & = \widehat K_i^{(2)} +  \left[\widehat K_i^{(1)},\ln|z_{12}|\right] \left(\beta_0+\frac12 \mathbf H^{(1)}\right),
%\delta K_L^{(2)}\,.
%
\label{HatNoHat}
\end{align}
etc.

Expanding the entries in the second equation in \eqref{spectrum-KN} to second order in $a_s$ and taking into account Eq.~ \eqref{reciprocity}
we obtain (for even $N$)
\begin{align}\label{Kcolorstructures}
   \widehat K_L^{(2)}(N) & = \beta_0 C_F \widehat K_L^{(2,\beta)}(N) +  C_F^2\widehat K_L^{(2,P)}(N)
    + \frac{C_F}{N_c} \widehat K_L^{(2,NP)}(N)
\end{align}
with
\begin{align}
\widehat K_L^{(2,\beta)}(N) & =  \frac{26}{3}\frac{1}{N(N+1)},
\notag\\
\widehat K_L^{(2,P)}(N) &= \frac{24}{N(N+1)}S_1 -\frac{130}3\frac1{N(N+1)},
\notag\\
\widehat K_L^{(2,NP)}(N) &=  - \frac{12+16S_{-2}}{(N-2)(N+3)}+\frac{16}{N(N+1)} \left(\frac56 - 3 \zeta_3 + S_3+S_1\right)
\notag\\&\quad
+\frac{32}{N(N+1)}\left(S_{1,-2}-\frac12S_{-3}\right) - \frac{16}{N(N+1)}\left(1+\frac2{N(N+1)}\right) S_{-2} \,.
\end{align}
Here $S_{\vec{a}}\equiv S_{\vec{a}}(N)$ are the harmonic sums~\cite{Vermaseren:1998uu}. The harmonic sums
$S_1$, $S_{-2}$,  $S_{3}$ and  $S_{1,-2}-\frac12 S_{-3}$ are parity invariant~\cite{Beccaria:2009vt}
so that the whole expression has this property.

Canonically invariant operators with the spectrum of eigenvalues corresponding to different terms
in these expressions are collected in Appendix~\ref{App:KernelList}. We obtain
\begin{align}
\widehat K_L^{(2,\beta)} &= \frac{26}{3} \mathcal H_+\,,
\notag\\
\widehat K_L^{(2,P)} &=   12  \mathcal H_{1,+} - \frac{130}{3} \mathcal H_+\,,
%\mathcal H_+ \left(12 \widehat{\mathcal H}  -\frac{130}{3} \right),
\notag\\
\widehat K_L^{(2,NP)} &=
-  8 \mathcal H_\text{sing}
+ 8 \left(\frac{5}{3} + \zeta_2 - 3 \zeta_3\right)  \mathcal H_+
+ 16 \mathcal H_+  \mathcal H_3
- 8 \mathcal H_+  \mathcal H_{-3} + 8 (1-\zeta_2)\mathcal H_{1,+}
\notag\\&\quad
- 8 \mathcal H_+   \mathcal H_{-2} - 16 \mathcal H_{++}  \mathcal H_{-2}   + 16 \zeta_2 \mathcal H_{++}\,,
\end{align}
and using \eqref{HatNoHat}
\begin{align}
K_L^{(2,\beta)} &= \widehat K_L^{(2,\beta)} + 4 \mathcal T_+\,,
\notag\\
K_L^{(2,P)} &= \widehat K_L^{(2,P)} +  8 \mathcal T_+ \left(  \widehat {\mathcal H} - \mathcal H_+ -\frac32 \right),
\notag\\
K_L^{(2,NP)} &= \widehat K_L^{(2,NP)},
\end{align}
where
\begin{align}
[\mathcal T_+ f](z_1,z_2)=\int_0^1d\alpha\int_0^{\bar\alpha} d\beta\, \ln(1-\alpha-\beta) f(z_{12}^\alpha,z_{21}^\beta)\,.
\end{align}

The longitudinal CF at the critical point in $4-2\epsilon_\ast$ dimensions in the $\overline{\text{MS}}$ scheme is given by
\begin{align}
C_{*,L}^{(2)}(\omega x, \omega) &= Y_2^{(0)}\otimes\Big( K_L^{(2)} +K_L^{(1)} \mathbb X^{(1)}\Big)
 +  Y_2^{(1)} \otimes \mathbf{H}^{(1)} K_L^{(1)}.
\end{align}
All convolutions can be taken using Eq.~\eqref{px}.

The analysis of the transverse CFs is  more complex but follows the same pattern. The expressions for the invariant kernel $\widehat
K_\perp^{(2)}$ are given in Appendix~\ref{app:Kperp}.
The remaining convolution integrals were done starting from the position-space expressions (as explained above) with the 
help of a Maple HyperInt package by E.~Panzer~\cite{Panzer:2014caa}.
The results are obtained in terms of generalized polylogarithms \cite{Goncharov:1998kja},
\begin{align}
\GPL(a_1,\ldots,a_n; x) &= \int_0^x\frac{\mathrm{d}t}{t-a_1}\GPL(a_2,\ldots,a_n;t)\qquad &&\text{if~}a_i\neq 0\text{~for~at~least~one~} i \in \mathbb{N}\,,\\
\GPL(\underbrace{0,\ldots,0}_{n~\text{times}}; x) &= \frac{1}{n!}\ln^n(x)\,\quad&&\text{for~} n\in \mathbb{N}_0\,,
\label{eq:gpldef}
\end{align}
in a filtration basis consisting of
\begin{alignat}{2}
&\GPL(a_1,\ldots,a_n; z)  \quad&&\text{with~}a_i \in \{\pm 1, \pm \omega \},\\
&\GPL(a_1,\ldots,a_n; \omega) \quad&&\text{with~}a_i \in \{\pm 1, 0\},
\end{alignat}
%$w \in \{0,\pm1, \pm \omega\}$.
%
with transcendental weight $n\leq 4$.
The final expressions for all 1-loop and 2-loop CFs contain 97 $\GPL$ functions and are provided in 
Mathematica format in the ancillary file {\tt CoefficientFunctions-FiltBasis.m} attached to this paper.
This representation is a convenient starting point for analytic continuations from the Euclidean
region to the different physical regions.
It has, however, the disadvantage that individual $\GPL$ functions contain spurious structures, as can be seen using symbol calculus~\cite{Goncharov:2010jf} for the transcendental functions.
Indeed, the symbols of the functions have letters $\{ \omega, \omega\pm 1,\omega \pm z, z \pm 1,   2 \}$, and all of
of them also appear as first entries, corresponding to logarithmic singularities.
The symbol of the sum entering the CFs has first entries $\{ \omega \pm 1, z\pm 1\}$, implying only singularities $\sim \ln^k(\omega \pm 1)$ and $\sim \ln^k(z \pm 1)$ in the results.
We note that the cancellation of spurious singular terms may lead to numerical instabilities near $\omega=0$ and $\omega=\pm z$ in this representation.
For the numerical evaluation of generalized polylogarithms, we have used the implementation of \cite{Vollinga:2004sn} in Ginac \cite{Bauer:2000cp}
and the {FastGPL} package~\cite{Wang:2021imw}.

To avoid spurious singularities and to reduce the number of transcendental functions which are difficult to numerically evaluate, we also consider alternative functional bases.
The Duhr-Gangl-Rhodes algorithm~\cite{Duhr:2011zq} allows us to construct a basis of functions with nested sums of low depth, which can significantly improve numerical evaluations, see e.g.~\cite{Bonciani:2013ywa,Gehrmann:2015ora}.
Here, we were able to map our results 
map our 1-loop and 2-loop results to a functional basis
consisting of 29  $\Li_{2,2}$, $\Li_{1,3}$ and $\Li_{1,2}$ functions plus classical polylogarithms and logarithms, 
all of which are free of spurious singularities and real-valued in the Euclidean region.
We provide our results in the ancillary file  {\tt CoefficientFunctions.m}, with all $\Li$ functions expressed in the $\GPL$ function notation (see \eqref{eq:lidef} for the $\Li$ function notation).
Using transformations of individual functions, we produced another representation free of spurious singularities.
While the resulting functions are more involved than in the previous case,
this result is the most compact representation of our results in the Euclidean region that we could find,
and we show it in Appendix~\ref{App:TwoLoopCFs}.
Unfortunately, analytic continuation to the physical region in Minkowski space becomes more tricky with these functions, and the prescription in \eqref{AnalyticContinuation} is only applicable for {\em real} values of the 
parameters $x$, $\eta$ and $\xi$.
These representations cannot be used, therefore, if the final convolution of the CF and the GPD is done 
using a deformed integration contour in the complex-$x$ plane.

We have verified that in the DVCS limit $\omega =1$ our results agree with~\cite{Braun:2020yib,Braun:2022bpn}.
One more check is provided by the structure of singular contributions $\sim \ln^k(1\pm z)/1\pm z)$ in the transverse CF,
see Appendix~\ref{App:ThresholdExpansion}. Our results agree with an independent calculation by J.~Schoenleber \cite{Schoenleber:2024dvq},
using threshold resummation techniques. Last but not least, we have verified that the two-loop CFs are analytic functions
in the limit when the ingoing and the outgoing photon momenta have the same absolute value but differ by sign, 
$q_1^2+q_2^2=0$. More precisely, the CFs are analytic functions in $\lambda$, defined by the  
rescaling $Q^2\mapsto \lambda Q^2$, $\omega\mapsto \omega/\lambda$, $z\mapsto z/\lambda$, in the limit 
$\lambda\to 0$. Analyticity for $\lambda\to 0$ is expected from the analysis of leading regions~\cite{Belitsky:2005qn},
which suggests that collinear factorization holds at the kinematic point $q_1^2 + q_2^2 = 0$ as well.

%%%%%%%%%%%%%%%%%%%%%%%%%%%%%%%%%%%%%%%%%%%%%%%%%%%%%%%%%%%%%%%%%%%%%%%%%%%%%%%%%%%%%%%%%%%%%%%%%%%%%%%%%%%%%%%%%%%%%%
\section{Numerical estimates}\label{sec:numerics}
%%%%%%%%%%%%%%%%%%%%%%%%%%%%%%%%%%%%%%%%%%%%%%%%%%%%%%%%%%%%%%%%%%%%%%%%%%%%%%%%%%%%%%%%%%%%%%%%%%%%%%%%%%%%%%%%%%%%%%

The numerical results in this section are presented for the invariant mass of the $\mu^+\mu^-$-pair
%final state photon virtuality
\begin{align}
   q_2^2 = 2.5~\text{GeV}^2\,, 
\end{align}
and two values
\begin{align}
  q_1^2 = - 0.6~\text{GeV}^2 \qquad\text{and}\qquad   q_1^2 = - 0.3~\text{GeV}^2\,,
\end{align}
which are considered realistic for the first DDVCS measurements 
in the JLAB12, JLAB20+ and EIC kinematics, see Ref.~\cite{Deja:2023ahc}.
The corresponding values of the $\omega$-parameter
\eqref{xi-eta-omega} are $\omega = -1.63158$ and $\omega = - 1.27273$.
The factorization scale is taken to be
\begin{align}  
 \mu^2 = \frac12 \left( q_2^2- q_1^2\right) 
\end{align}
and the value of the strong coupling (the same in both cases)
$\alpha_s = 0.4$  for three active flavors, $n_f=3$.
The CFs are continued analytically from the Euclidean region using the prescription in 
Eq.~\eqref{AnalyticContinuation}. 
Here, we make use of the expressions collected in the ancillary file {\tt CoefficientFunctions-FiltBasis.m}
and the {\tt FastGPL} C++ library \cite{Wang:2021imw} for the numerical evaluation of generalized polylogarithms. 

We employ the toy GPD model from Ref.~\cite[Eq.~(3.331)]{Belitsky:2005qn}
in order to estimate the size of the NNLO correction to the Compton form factor 
$\mathcal F_\perp(\xi,\eta)$ \eqref{CFFfactorization}. It is based
on the so-called double-distributions ansatz \cite{Radyushkin:1997ki} and allows for a simple analytic representation:
\begin{align}
H(x,\xi)&=\frac{(1-n/4)}{\xi^3}\biggl[\theta(x+\xi)\left(\frac{x+\xi}{1+\xi}\right)^{2-n}\Big(\xi^2-x+(2-n)\xi(1-x)\Big)
-(\xi\to -\xi)\biggr]\, .
\label{GPDmodel}
\end{align}
%
%%%%%%%%%%%%%%%%%%%%%%%%%%%%%%%%%%%%%%%%%%%%%%%%%%%%%%%%%%%%%%%%%%%%%%%%%%%%%%%%%%
An overall normalization is irrelevant for our purposes so we omit it.
We use the value of the parameter $n=1/2$ which corresponds to a valence-like PDF $q(x)\sim x^{-1/2} (1-x)^3$ in the forward limit.
%The $C$-even part of the GPD \eqref{GPDmodel}, $H(x,\xi)-H(-x,\xi)$, 
%is shown in Fig.~\ref{fig:HGPD} for several values of $\xi$.

\begin{figure}[t]
\begin{center}
\includegraphics[width=0.62\linewidth]{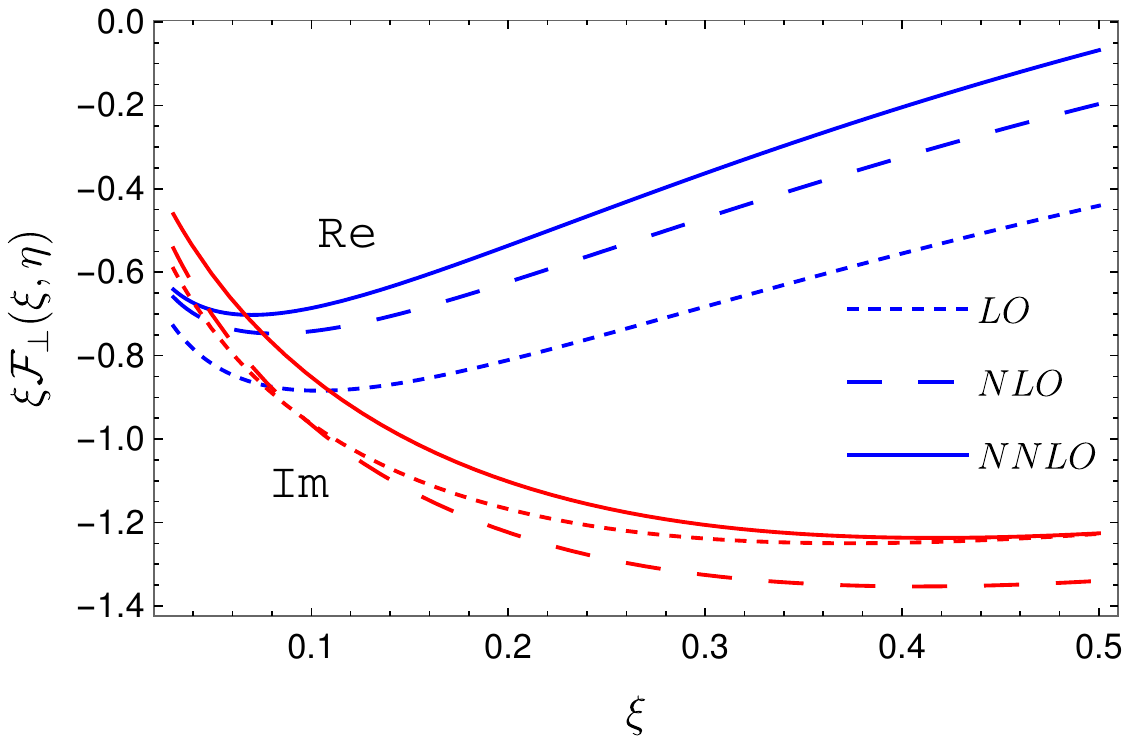}\\
\includegraphics[width=0.62\linewidth]{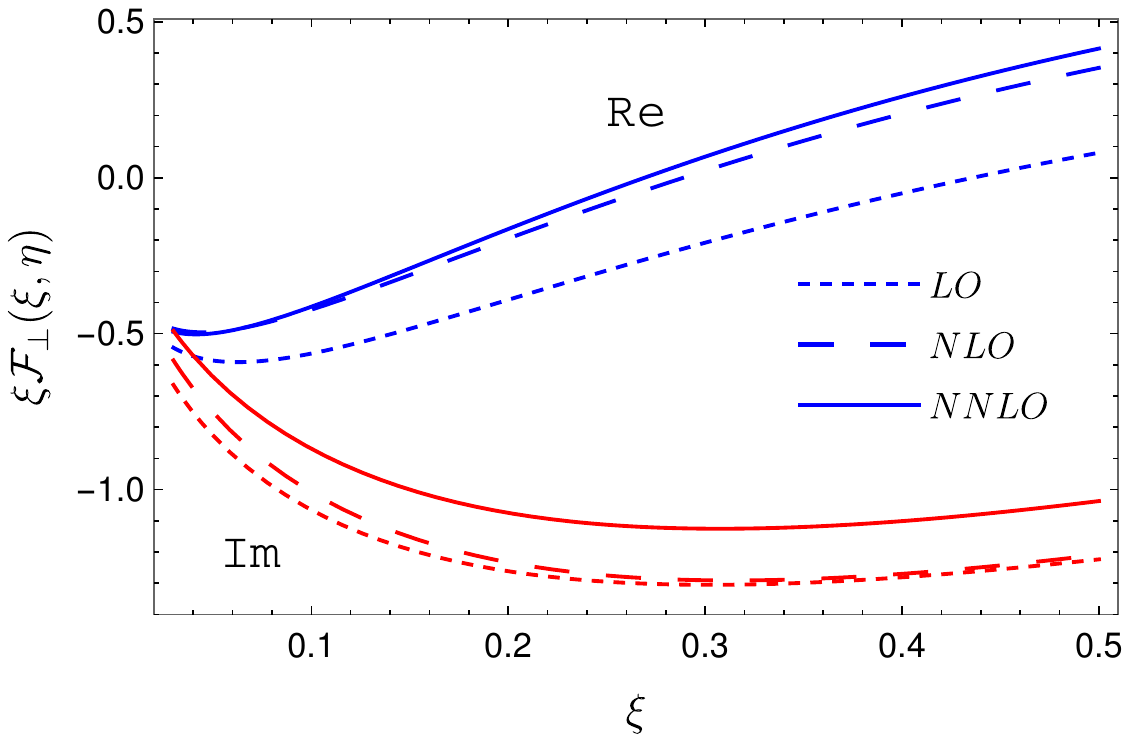}
\end{center}
\caption{Real (blue) and imaginary (red) parts of the CFF $\mathcal F_\perp(\xi,\eta)$ \eqref{CFFfactorization}
as a function of $\xi$ for $\xi/\eta = -1.63158$ (upper panel) and  $\xi/\eta = - 1.27273$ (lower panel)
for the GPD model in Eq.~\eqref{GPDmodel}.
The leading-order results, and the results including one-loop and two-loop corrections
are shown by the short dashes, long dashes and solid curves, respectively. 
%Normalization corresponds to Eq.~\eqref{GPDmodel}.
}
\label{fig:CFFT}
\end{figure}

\begin{figure}[t]
\begin{center}
\includegraphics[width=0.62\linewidth]{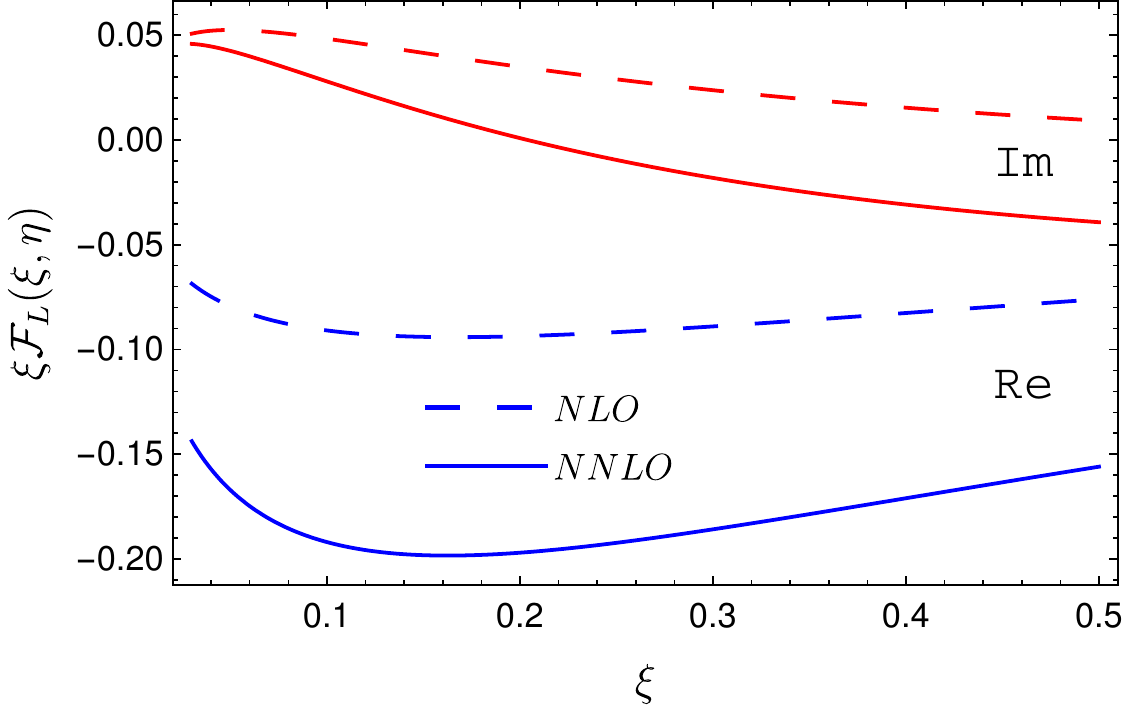}\\
\includegraphics[width=0.62\linewidth]{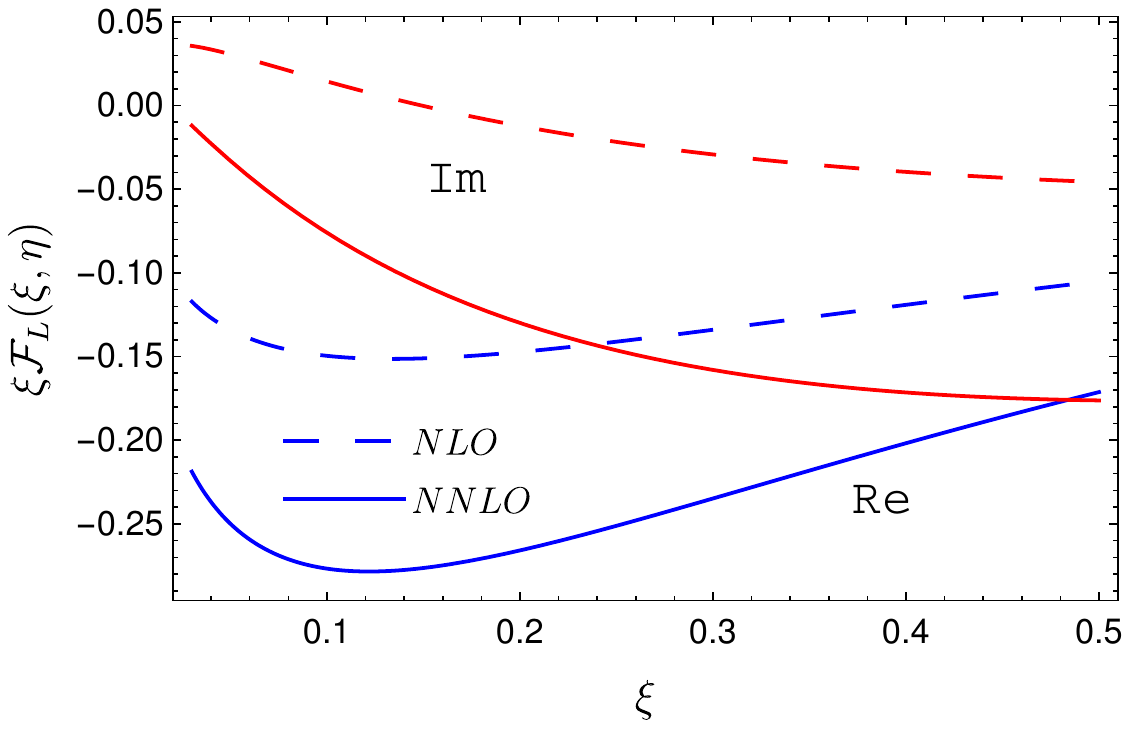}
\end{center}
\caption{Real (blue) and imaginary (red) parts of the CFF $\mathcal F_L(\xi,\eta)$ \eqref{CFFfactorization}
as a function of $\xi$ for $\xi/\eta = -1.63158$ (upper panel) and  $\xi/\eta = - 1.27273$ (lower panel)
for the GPD model in Eq.~\eqref{GPDmodel}
The leading-order results, and the results including both one-loop and two-loop corrections
are shown by the long dashes and solid curves, respectively. 
%Normalization corresponds to Eq.~\eqref{GPDmodel}.
}
\label{fig:CFFL}
\end{figure}

For a numerical evaluation of the convolution integrals in the $x<\xi$  region
it proves to be convenient to shift the integration contour to the complex plane. We have checked that the
results do not depend on the shape of the integration contour, which is a good test of numerical accuracy.
The results for the transverse and the longitudinal CFFs \eqref{CFFfactorization} 
are  shown in Fig.~\ref{fig:CFFT} and Fig.~\ref{fig:CFFL}, respectively. 
We show real (blue) and imaginary (red) parts of the CFFs
as a function of $\xi$ for the fixed value of $\omega = \xi/\eta = -1.63158$ (upper panels) and  
$\omega = \xi/\eta = - 1.27273$ (lower panels).
The leading-order (LO) results are shown by short dashes, and the calculation 
including one-loop (NLO) and two-loop corrections (NNLO) by the long dashes and solid curves, 
respectively. 

One sees that the corrections are in general quite large (for the chosen kinematics) and 
have a nontrivial structure. In particular for $\mathcal F_\perp(\xi,\eta)$, the NLO (one loop) 
corrections are large for the real part and small for imaginary part of the CFF, whereas
the NNLO (two-loop) corrections, on the contrary, are small for the real part and large
for imaginary part.   The NNLO corrections for  $\mathcal F_L(\xi,\eta)$ are very large so that 
the perturbative expansion does not show any sign of convergence for this case. These features certainly 
call for an increase of the invariant mass of the lepton pair which, hopefully, will become 
possible in future experiments.

As far as the relative contributions of the three color structures \eqref{color} in the NNLO correction
are concerned, the terms proportional to the QCD $\beta$-function prove to be the largest, but are 
partially compensated by contributions of ``planar'' diagrams $\sim C_F^2$. 
The non-planar contributions $\sim C_F/N_c$ are in all cases an order of magnitude below the planar 
ones.

%%%%%%%%%%%%%%%%%%%%%%%%%%%%%%%%%%%%%%%%%%%%%%%%%%%%%%%%%%%%%%%%%%%%%%%%%%%%%%%%%%%%%%%%%%%%%%%%%%%%%%%%%%%%%%%%%%%%%%%%%%%%%%%%%%%%%

%%%%%%%%%%%%%%%%%%%%%%%%%%%%%%%%%%%%%%%%%%%%%%%%%%%%%%%%%%%%%%%%%%%%%%%%%%%%%%%%%%%%%%%%%%%%%%%%%%%%%%%%%%%%%%%%%%%%%%
\section{Summary }\label{sec:summary}
%%%%%%%%%%%%%%%%%%%%%%%%%%%%%%%%%%%%%%%%%%%%%%%%%%%%%%%%%%%%%%%%%%%%%%%%%%%%%%%%%%%%%%%%%%%%%%%%%%%%%%%%%%%%%%%%%%%%%%

Using the approach based on conformal symmetry~\cite{Braun:2013tva,Braun:2020yib} we have calculated the two-loop coefficient functions in double deeply virtual Compton scattering
in the $\overline{\text{MS}}$ scheme for the flavor-nonsinglet vector contributions.
Analytic expressions for the coefficient functions in momentum fraction space are
presented in Appendix~\ref{App:TwoLoopCFs} and in two ancillary files using different representations for the 
relevant generalized polylogarithms. 
Numerical estimates in Sect.~\ref{sec:numerics} suggest that the two-loop contribution to the Compton form factors
at the scale of proposed experiments is significant.

The technique developed in this work can be used to calculate the two-loop contributions to the 
flavor-nonsinglet coefficient functions for the correlation functions of all quark-antiquark 
currents with applications to, e.g., the light-cone sum rules for the pion electromagnetic and transition
form factors \cite{Braun:1999uj,Agaev:2010aq}.
%The corresponding calculations go beyond the tasks of this study.

%%%%%%%%%%%%%%%%%%%%%%%%%%%%%%%%%%%%%%%%%%%%%%%%%%%%%%%%%%%%%%%%%%%%%%%%%%%%%%%%%%%%%%%%%%%%%%%%%%%%%%%%%%%%%%%%%%%%%%
\section*{Acknowledgments}
%%%%%%%%%%%%%%%%%%%%%%%%%%%%%%%%%%%%%%%%%%%%%%%%%%%%%%%%%%%%%%%%%%%%%%%%%%%%%%%%%%%%%%%%%%%%%%%%%%%%%%%%%%%%%%%%%%%%%%
\addcontentsline{toc}{section}{Acknowledgments}

We thank J.~Wagner and L.~Szymanowski for the discussion of analytic continuation properties of the DDVCS amplitudes.   
This study was supported by Deutsche Forschungsgemeinschaft (DFG) through the Research Unit FOR 2926, ``Next Generation pQCD for
Hadron Structure: Preparing for the EIC'', project number 40824754. In addition, 
H.-Y.J. gratefully acknowledges support from the National Natural Science Foundation of China with Grant No. 12405114.

%%%%%%%%%%%%%%%%%%%%%%%%%%%%%%%%%%%%%%%%%%%%%%%%%%%%%%%%%%%%%%%%%%%%%%%%%%%%%%%%%%%%%%%%%%%%%%%%%%%%%%%%%%%%%%%%%%%%%%
%%%%%%%%%%%%%%%%%%%%%%%%%%%%%%%%%%%%%%%%%%%%%%%%%%%%%%%%%%%%%%%%%%%%%%%%%%%%%%%%%%%%%%%%%%%%%%%%%%%%%%%%%%%%%%%%%%%%%%
%%%%%%%%%%%%%%%%%%%%%%%%%%%%%%%%%%%%%%%%%%%%%%%%%%%%%%%%%%%%%%%%%%%%%%%%%%%%%%%%%%%%%%%%%%%%%%%%%%%%%%%%%%%%%%%%%%%%%%
%%%%%%%%%%%%%%%%%%%%%%%%%%%%%%%%%%%%%%%%%%%%%%%%%%%%%%%%%%%%%%%%%%%%%%%%%%%%%%%%%%%%%%%%%%%%%%%%%%%%%%%%%%%%%%%%%%%%%%

\appendix
\addcontentsline{toc}{section}{Appendices}
\renewcommand{\theequation}{\Alph{section}.\arabic{equation}}
\renewcommand{\thesection}{{\Alph{section}}}
\renewcommand{\thetable}{\Alph{table}}
\setcounter{section}{0} \setcounter{table}{0}
%####################################################################################################################################
\section*{Appendices}
%####################################################################################################################################

\section{Helicity amplitudes}\label{Appendix:A}

In this Appendix, we discuss the decomposition of the generalized Compton amplitude in terms of helicity amplitudes.
It is convenient \cite{Braun:2012bg} to use the photon momenta $q_1$, $q_2$ to define the longitudinal plane
spanned by two light-like vectors $n^\mu$, $\tilde n^\mu$.
In the present context we can put $\Delta^2=0$ and define
\begin{align}
  n^\mu = \frac{q_1^\mu}{q_1^2} - \frac{q_2^\mu}{q_2^2}\,, \qquad   \tilde n^\mu = q_1^\mu-q_2^\mu = \Delta^\mu\,, \qquad (n\tilde n) = \frac12 \frac{(q_1^2-q_2^2)^2}{q_1^2q_2^2}
= \frac{2\omega^2}{1-\omega^2}.
\end{align}
The amplitude \eqref{Compton} can be expanded in terms of helicity amplitudes to twist-two accuracy as follows:
\begin{align}
A^{\text{t2}}_{\mu\nu}&=\epsilon_\mu^+ \epsilon_\nu^- A^{+-} +\epsilon_\mu^- \epsilon_\nu^+ A^{-+}  +\epsilon_\mu^L(q) \epsilon_\nu^L(q') A^L
=-g_{\mu\nu}^\perp A_V +\epsilon_{\mu\nu}^\perp A_A +\hat\epsilon_\mu^L(q_1) \hat\epsilon_\nu^L(q_2) A_L\,.
\end{align}
Here
\begin{align}
g_{\mu\nu}^\perp = g_{\mu\nu} -\frac{n^\mu \tilde n^\nu + n^\nu \tilde n^\mu}{(n\tilde n)}, \qquad\quad
\epsilon_{\mu\nu}^\perp =\frac1{(n\tilde n)}\epsilon_{\mu\nu\alpha\beta}n^\alpha \tilde n^\beta,
\end{align}
$\epsilon^\pm_\mu$ are orthogonal unit vectors in the transverse plane that can be taken as transverse polarization vectors for both initial and final photons,
and the longitudinal photon polarization vectors are given by
\begin{align}
\hat \epsilon_L^\mu(q_1) & = \frac{1}{wQ\sqrt{1+\omega}} \Big[q_1^\mu- q_2^\mu(1+\omega)\Big]\,,
\notag\\
\hat \epsilon_L^\nu (q_2) & =\frac{1}{wQ\sqrt{1-\omega}} \Big[q_2^\nu - q_1^\nu(1-\omega)\Big]\,.
\end{align}
The longitudinal helicity amplitude can therefore be projected as
\begin{align}
  A_L &=
% \frac1{Q^4(1-\omega^2) \omega^4}
\hat \epsilon_L^\mu(q_1) T_{\mu\nu} \hat \epsilon_L^\nu(q_2)
=  - \frac{\sqrt{1-\omega^2}}{\omega^2 Q^2} \Delta^\mu T_{\mu\nu} \Delta^\nu,
\end{align}
where we used that, since $q_1^\mu  T_{\mu\nu} =  T_{\mu\nu} q_2^\nu =0$, one can replace
\begin{align}
\hat \epsilon_L^\mu(q_1) \mapsto \frac{\sqrt{1+\omega}}{\omega Q}\Delta^\mu,
\qquad\quad
\hat \epsilon_L^\nu(q_2) \mapsto - \frac{\sqrt{1-\omega}}{\omega Q} \Delta^\nu.
\end{align}
Comparing these expressions with the the conventional decomposition in terms of (generalized) Compton form factors in \eqref{F12L}, we get
\begin{align}
   A_V &= \mathcal{F}_1 = \mathcal{F}_\perp\,, \qquad\quad A_L = - \sqrt{1-\omega^2} \mathcal{F}_L\,.
\end{align}
Note that the longitudinal CFF $\mathcal{F}_L$ does not vanish for $\omega =\pm 1$, but it does not contribute to
DVCS and TCS thanks to the $\sqrt{1-\omega^2}$ prefactor.

\allowdisplaybreaks

%%%%%%%%%%%%%%%%%%%%%%%%%%%%%%%%%%%%%%%%%%%%%%%%%%%%%%%%%%%%%%%%%%%%%%%%%%%%%%%%%%%%%%%%%%%%%%%%%%%%%%%%%%%%%%%%%%%%%%%%%%%%%%%%%%%%%%%
\section{\texorpdfstring
{The $\widehat{\mathcal K}_\perp$ kernel}
{The Khat-perp kernel}
}\label{app:Kperp}
%%%%%%%%%%%%%%%%%%%%%%%%%%%%%%%%%%%%%%%%%%%%%%%%%%%%%%%%%%%%%%%%%%%%%%%%%%%%%%%%%%%%%%%%%%%%%%%%%%%%%%%%%%%%%%%%%%%%%%%%%%%%%%%%%%%%%%%
Here, we present the eigenvalues of the $\widehat K_\perp$ kernels employed in Sect.~\ref{sec:twoloopCFs}:
\begin{align}
&\widehat{\mathcal K}_\perp^{(2,\beta)}(N) =\left(2\zeta_2+\frac59\right)S_1 -\left(\zeta_2+\frac{10}9\right) \frac1{N(N+1)} %-\frac{5}{N^2(N+1)^2}
+2\zeta_3-\frac{65}6 \zeta_2+\frac{45}8,
\notag\\[2mm]
&\widehat{\mathcal K}^{(2,P)}_{\perp}(N) =
\notag\\ &=
\frac12\left( \mathcal K_\perp^{(1)}(N)\right)^2+4\zeta_2 (\bar\gamma^{(1)}_N)^2
+ 4\zeta_3 \left(11+\frac{12}{N(N+1)}\right)
-64\zeta_3S_1 -8\zeta_2 S_1^2 +\frac{6S_{-2}}{N(N+1)}
\notag\\
&\quad
+\frac{{2S_1^2}}{N(N+1)}
+12\zeta_2
 \left(-{1}+\frac{{2}}{3N(N+1)}\right)S_1
+\left(\frac{149}9 -\frac{8}{N(N+1)}-\frac{{2}}{N^2(N+1)^2}
\right)S_1
\notag\\
&\quad
+\frac{11}8 +\frac{11}3\zeta_2 +8\zeta_2^2-\frac{19}{9N(N+1)}-\frac{{16}}{N^2(N+1)^2}-\frac{{2}}{N^3(N+1)^3}\,,
\notag \\[2mm]
&\widehat{\mathcal K}^{(2,NP)}_\perp(N) =
\notag\\&=
-12 S_{-2}^2-8S_{-4}
+ 4\Big(2S_{1,3}- S_4\Big)
-\frac{12 S_3}{N (N+1)} +\frac{24 \left(S_{-3}-2 S_{1,-2}\right)}{N (N+1)}+\frac{16 S_1S_{-2}}{N(N+1)}
\notag\\
&\quad
+\left(
\frac{36}{N^2 (N+1)^2}
+\frac{24}{(N-2) (N+3)}+\frac{52}{N (N+1)}+8\right)S_{-2}
   \notag\\
&\quad
+\left(-\frac{8}{N^3 (N+1)^3}-\frac{8}{N^2(N+1)^2}-\frac{{10}}{ N (N+1)}+{\frac{70}{9}} \right) S_1
 \notag\\
&\quad
-\frac{{68}}{9N(N+1)}
+\frac{18}{(N-2) (N+3)}-\frac{35}{4} +\left(\frac{50}{N (N+1)}+54\right) \zeta_3
\notag  \\
&\quad
- 4 \zeta_2^2
-36 \zeta_3 S_1  -\frac{12 \zeta_2 S_1}{N (N+1)}
+ \zeta_2 \left(\frac{8}{N^2 (N+1)^2}+\frac{4}{N (N+1)}-\frac{20}{3}\right)\,.
%%;
\end{align}

%%%%%%%%%%%%%%%%%%%%%%%%%%%%%%%%%%%%%%%%%%%%%%%%%%%%%%%%%%%%%%%%%%%%%%%%%%%%%%%%%%%%%%%%%%%%%%%%%%%%%%%%%%%%%%%%%%%%%%%%%%%%%%%%%%%%%%%
\section{\texorpdfstring
{$SL(2)$-invariant kernels}
{SL(2)-invariant kernels}
}\label{App:KernelList}
%%%%%%%%%%%%%%%%%%%%%%%%%%%%%%%%%%%%%%%%%%%%%%%%%%%%%%%%%%%%%%%%%%%%%%%%%%%%%%%%%%%%%%%%%%%%%%%%%%%%%%%%%%%%%%%%%%%%%%%%%%%%%%%%%%%%%%%

We collect here the invariant kernels and their eigenvalues used in Sect.~\ref{sec:twoloopCFs}.
Let
\begin{align}
\mathrm M_n[\omega]=\int_0^1 d\alpha\int_0^{\bar\alpha} \!d\beta\,\omega(\tau) (1-\alpha-\beta)^{n-1}\,,
\qquad \tau =\frac{\alpha\beta}{\bar\alpha\bar\beta}\,,
\qquad \bar\tau = 1-\tau\,,
\end{align}
where $n$ is even. One obtains
\begin{align}
\mathcal H_{+}&: \qquad \mathrm M_n[1] = \frac{1}{n(n+1)}\,,
\qquad
&& \mathcal H_{++}: \qquad \mathrm M_n[-\ln\bar \tau] = \frac{1}{n^2(n+1)^2}\,,
\notag\\
\mathcal{H}_{1,+}&: \qquad \mathrm M_n[-\ln\tau] =\frac{2 S_1(n)}{n(n+1)}\,,
\qquad
&&  \mathcal{H}_{-2}: \qquad \mathrm M_n[\bar\tau]= 2S_{-2}(n)+\zeta_2\,,
\notag\\
\mathcal{H}_{-2,+}&: \qquad \mathrm M_n[\Li_2(\tau)]=\frac{2S_{-2}(n)+\zeta_2}{n(n+1)}\,,
\qquad
&& \mathcal{H}_3: \qquad  \mathrm M_n\left[\frac{\bar\tau}{2\tau}\ln\bar\tau\right]= S_3(n)-\zeta_3\,,
\end{align}
and
\begin{align}
&\mathcal{H}_{-3}: \qquad
\mathrm M_n\left[-\bar\tau\ln\bar\tau\right] = 2S_{-3}(n)-4S_{1,-2}(n)-2\zeta_2 S_1(n)+ \zeta_3\,,
\notag\\
&\mathcal{H}_{-4}: \qquad
\mathrm M_n\left[\bar\tau\left(\Li_2(\tau)+\frac12\ln^2\bar\tau\right)\right]
= 2S_{-4}(n)+\frac7{10}\zeta_2^2\,,
\notag\\
&\mathcal{H}_{1,3}: \qquad
\mathrm M_n\left[\frac{\bar\tau}{4\tau}\left(\Li_2(\tau)+\frac12\ln^2\bar\tau\right)\right]
= S_{1,3}(n)-\frac12 S_4(n) -\zeta_3 S_1(n) +\frac3{10}\zeta_2^2\,,
\notag\\
&\mathcal{H}_{\mathrm sing}: \qquad
\mathrm M_n\left[
\frac{\tau^2\!+\!4\tau\! +\!1}{(\tau-1)^2}\big[ \Li_2(\tau)-\zeta_2\big] + 3 \Big(\frac{\tau\!+\!1}{\tau\!-\!1}\Big) \ln(1\!-\!\tau) -
\frac32\frac{3\tau+1}{(\tau-1)}\right]
= \frac{2S_{-2}(n) + 3/2}{(n-2)(n+3)}.
\end{align}

%%%%%%%%%%%%%%%%%%%%%%%%%%%%%%%%%%%%%%%%%%%%%%%%%%%%%%%%%%%%%%%%%%%%%%%%%%%%%%%%%%%%%%%%%%%%%%%%%%%%%%%%%%%%%%%%%%%%%%%%%%%%%%%%%%%%%%%
\section{Restoring the scale dependence}\label{App:ScaleDependence}
%%%%%%%%%%%%%%%%%%%%%%%%%%%%%%%%%%%%%%%%%%%%%%%%%%%%%%%%%%%%%%%%%%%%%%%%%%%%%%%%%%%%%%%%%%%%%%%%%%%%%%%%%%%%%%%%%%%%%%%%%%%%%%%%%%%%%%%

In this Appendix, we provide details for the restoration of the scale dependence of the coefficient functions
mentioned in Sect.~\ref{sec:solansatzinvkernel}.
The scale-dependent terms $\sim \ln Q/\mu, \ln^2 Q/\mu$  in the CFs are completely fixed by the renormalization group equations.
Since the evolution kernel in the $\overline{\rm MS}$ scheme does not depend on $\epsilon$,
$\mathbb{H}(a_s,\epsilon) = \mathbb{H}(a_s)$, in a generic $d$-dimensional theory
\begin{align}
\Big(\mu\partial_\mu + \beta(a_s,\epsilon)\partial_{a_s}\Big)C\left(Q^2/\mu^2,a_s,\epsilon\right) =
C\left(Q^2/\mu^2,a_s,\epsilon\right)\otimes  \mathbb H(a_s)\,,
\end{align}
where
\begin{align}
    C\otimes\mathbb H = \int_{-1}^1 dx'\, C(x')\, \mathbb H(x',x)\,.
\end{align}
Solving this equation one obtains \cite{Braun:2020yib}
\begin{align}
C(\sigma,a_s,\epsilon) & =\Big( C^{(0)}+ a_s C^{(1)}(\epsilon)+a_s^2 C^{(2)}(\epsilon)+\ldots \Big)\otimes
\Big(1 -\frac12 \ln \sigma\, \mathbb H(a_s)+ \frac18 \ln^2\!\sigma\, \mathbb H^2(a_s) +\ldots   \Big)\,
\notag\\
&\quad
-\beta(a_s,\epsilon)\left( -\frac12 C_1(\epsilon)\,\ln \sigma + \frac1{8a_s} \ln^2\sigma\,C_0\otimes \mathbb H(a_s)\right)
+ O(a_s^3,a_s^2\epsilon, a_s\epsilon^2)\,.
\end{align}
Here $\sigma = Q^2/\mu^2$ and $C^{(0)}, C^{(1)}(\epsilon), C^{(2)}(\epsilon)$ are the CFs in $d$ dimensions \eqref{eq:generic-d}
at $\mu^2=Q^2$,
alias $\sigma=1$. Note that the contribution in the second line vanishes at the critical point, $\beta(a_s,\epsilon_\ast)=0$.
For the physical case $d=4$ one obtains
\begin{align}
C(\sigma,a_s,\epsilon=0) & = C^{(0)}+ a_s \biggl(C^{(1)}_\ast -\frac12 \ln \sigma \, C^{(0)}\otimes \mathbb H^{(1)}\biggr)
+a_s^2\biggl\{C_2^{(*)}+\beta_0 C^{(1,1)}
\notag\\&\quad
-\frac12 \ln \sigma \biggl[ C^{(0)}\!\otimes\mathbb H^{(2)}
+2 C^{(1)}\!\otimes \Big(\beta_0+\frac12\mathbb H^{(1)}\Big) \biggr]
\notag\\&\quad
+\frac14 \ln^2\!\sigma\, C^{(0)}\!\otimes \mathbb H^{(1)}\Big(\beta_0+\frac12\mathbb H^{(1)} \Big)\!\biggr\}\,,
\end{align}
where the CFs in $d=4$ are related to the ones at the critical point as
$C^{(1)}(\epsilon = 0)= C^{(1)}_\ast$ and $C^{(2)}(\epsilon = 0)= C_\ast^{(2)}+\beta_0 C^{(1,1)}$, see Eq.~\eqref{eq:d=4-restored}.

%%%%%%%%%%%%%%%%%%%%%%%%%%%%%%%%%%%%%%%%%%%%%%%%%%%%%%%%%%%%%%%%%%%%%%%%%%%%%%%%%%%%%%%%%%%%%%%%%%%%%%%%%%%%%%%%%%%%%%%%%%%%%%%%%%%%%%%
\section{Two-loop coefficient functions}\label{App:TwoLoopCFs}
%%%%%%%%%%%%%%%%%%%%%%%%%%%%%%%%%%%%%%%%%%%%%%%%%%%%%%%%%%%%%%%%%%%%%%%%%%%%%%%%%%%%%%%%%%%%%%%%%%%%%%%%%%%%%%%%%%%%%%%%%%%%%%%%%%%%%%%

In this Appendix, we present explicit results for the two-loop coefficient functions discussed in Sect.~ \ref{sec:twoloopCFs}.
We express our results in terms of multiple polylogarithms
\begin{align}
\Li_{n_1,\ldots,n_r}(z_1,\ldots,z_r) = \sum_{0<k_1<\ldots < k_r} \frac{z_1^{k_1}\cdots z_r^{k_r}}{k_1^{n_1}\cdots k_r^{n_r}}\,,
\label{eq:lidef}
\end{align}
where the depth $r$ denotes the number of summations.
We note that this convention follows those of Refs.~\cite{Goncharov:1998kja,Borwein:1999js} and the HyperInt package~\cite{Panzer:2014caa}, whereas the order of subscripts and arguments need to be reversed to match the conventions of Ginac's $\Li$ functions~\cite{Vollinga:2004sn}.
The $\Li$ functions can easily be converted into $\GPL$ function notation \eqref{eq:gpldef} and vice versa.
Threshold expansions of our results are presented in Appendix \ref{App:ThresholdExpansion} below.

%%%%%%%%%%%%%%%%%%%%%%%%%%%%%%%%%%%%%%%%%%%%%%%%%%%%%%%%%%%%%%%%%%%%%%%%%%%%%%%%%%%%%%%%%%%%%%%%%%%%%%%%%%%%%%%%%%%%%%%%%%%%%%%%%%%%%%%

                          \subsection{Longitudinal CF}\label{App:TwoLoopLong}

%%%%%%%%%%%%%%%%%%%%%%%%%%%%%%%%%%%%%%%%%%%%%%%%%%%%%%%%%%%%%%%%%%%%%%%%%%%%%%%%%%%%%%%%%%%%%%%%%%%%%%%%%%%%%%%%%%%%%%%%%%%%%%%%%%%%%%%

As was discussed in Sect.~\ref{sec:framework}, the difference between the critical and four dimensional CFs at two loops,
$C_{L}^{(2)}(z,\omega) - C_{*,L}^{(2)}(z,\omega) = \beta C_L^{(11)}(z,\omega)$ comes from the $\epsilon$-expansion of the one-loop CF in
$d=4-2\epsilon$ dimensions. Taking this contribution into account and adding the RG logarithms as explained in 
App.~\ref{App:ScaleDependence},
we write the two-loop longitudinal CF using the notations in~Eqs.~\eqref{loopCFs},\eqref{color},\eqref{RGlogs}.
The $A_L$ -type contributions all vanish. 
The terms $\sim \ln(Q^2/\mu^2)$ in the $B_L$-functions 
corresponding to the different color structures, Eq.~\eqref{color}, take the following form:
\begin{align}
B_{L,\ln}^{(2,\beta)}(z,w) & = -2 \mathrm L_1(z,w)\,,
\notag\\
B_{L,\ln}^{(2,P)}(z,w) & = \mathrm L_2(z,-w) - \mathrm L_2(w,-w) +\mathrm L_1(z,w)\left(\mathrm L_1(z,-w)+\frac32\right),
\notag\\
B_{L,\ln}^{(2,NP)}(z,w) & =0\,.
\end{align}
The remaining contributions are:

\noindent $\bullet$~~Terms $\sim \beta_0 C_F$:
\begin{flalign}\label{BLbeta}
B_L^{(2,\beta)}(z,w) & = 2\,\mathrm L_2(z,w) + \frac{13}3\mathrm L_1(z,w)
+\Big[ -\ln^2(1-z) +\ln^2(1-w) +2\mathrm L_1(z,w) \Big]\,.
\end{flalign}
The terms in square brackets in Eq.~\eqref{BLbeta} originate from $C^{(11)}$.\\[2mm]

\noindent $\bullet$~~Planar contributions $\sim C_F^2$:
\allowdisplaybreaks{
\begin{flalign} \label{BLP}
B_L^{(2,P)}(z,w)&=4\biggl\{
\mathrm L_{12}(z,w) + \mathrm L_3(z,-w) - \mathrm L_3(w,-w)  + \Big( \mathrm L_2(z,-w) -\mathrm L_2(w,-w)\Big)\ln(1 + w)
&
\notag\\
&\quad
+ \frac13\Big(\ln^3(1 - z) -\ln^3 (1-w)\Big) - \frac32\mathrm L_2(z,w) %- \mathrm L_2(z,-w)\ln(1 + w)
+ \frac 14 \Big(\ln^2(1-z) - \ln^2(1-w) \Big) &
\notag\\
&\quad
    - \frac12\left( \ln^2(1+w) + \ln^2(1 - w)  - \ln(1-w^2)  +\frac{ 47}{6} \right)\mathrm L_1(z,w)
    \biggr\}. &
\end{flalign}

\noindent $\bullet$~~Non-planar contributions $\sim C_F/N_c$ are considerably more complicated:
\begin{flalign}
\label{BLNP}
&B_L^{(2,NP)}(z,w)~=
\notag\\&\quad=
4\biggl \{  \mathrm L_{2}(z,w)\Big[ \mathrm L_{2}(z,-w) + \mathrm L_{2}(-z,w)\Big]
+ \widehat{\,\mathrm M}_{22}(-z,-w) - \widehat{\,\mathrm M}_{22}(-w,-z)  + 6\, \mathrm L_{112}(z,w) &
\notag\\&\qquad\quad
+ \mathrm L_1(z,w)
\left[ \widehat{L}_3(z,w) +4 \mathrm L_{12}(z,w)-6\zeta_3\right]
-\frac12\Big[\mathrm L_1^2(z,-w)-\mathrm L_1^2(w,-w)\Big]\mathrm L_2(-z,w) &
\notag\\&\qquad\quad
+\mathrm L_2(z,w)\Big[\mathrm L^2_1(z,w) + \frac12 \mathrm L_1^2(w,-w)\Big]
 +  \mathrm L_1(-z,-w) \mathrm L_2(z,w)
 +  \mathrm L_1(-z,w) \mathrm L_2(z,-w) &
 \notag\\&\qquad\quad
 - \frac16 \mathrm L^3_1(w,-w) \mathrm L_1(z,w)
  - \frac14 \mathrm L^2_1(w,-w) \mathrm L_1(z,-z)
  - \mathrm L_1(z,-z)\mathrm L_1(z,w) + \frac53 \mathrm L_1(z,w)
  \biggr\}   &
\notag\\&\qquad\quad
  - \frac{2(w^2-3)}{w^2}
 	\left\{
\mathrm L_1(-z,-w) \mathrm L_2(z,w) +\mathrm L_1(-z,w) \mathrm L_2(z,-w) -\frac14\mathrm L^2_1(w,-w)\mathrm L_1(z,-z)
 	 \right\}
&\notag\\&\quad\quad
  - \frac{12}{w}\left\{ \mathrm L_2(z,-w) + \frac14 \mathrm L^2_1(w,-w)  -\frac12(w-2)\mathrm L_1(z,w)\right\}
&\notag\\&\quad\quad
+6 (w-z) \Biggl\{ \frac{z}{w^4}(w^2-3)
\biggl[ \widehat{ D}_L(w) -\mathrm M_{22}(z,w) 
	+\frac12\mathrm L_{2}(z,w)\left(\mathrm L_{2}(-z,w)  +\frac12 \mathrm L^2_1(w,-w)\right)
&\notag\\&\qquad\qquad\qquad
-\mathrm L_4(-z,-w) +\mathrm L_2(z,w)\mathrm L_2(-1,-w) -\frac14\mathrm L_1(w,-w)\left[\widehat{L}_3(z,w) +6\zeta_3\right]
&\notag\\&\qquad\qquad\qquad
-\frac{2 w} z\biggl(
	\mathrm L_1(z,-w) \mathrm L_2(-z,w)  + \!\frac14 \mathrm L^2_1(w,-w)\mathrm L_1(z,w)
\biggr)
\biggr]
&\notag\\&\qquad\qquad\qquad
+\frac{6z}{w^3} \biggl[
	\frac14  \widehat{L}_3(z,w) -\frac12\widehat{L}_{12}(z,w) - \frac32\zeta_3
	-\left(\frac14 \mathrm L^2_1(w,-w)  +  \mathrm L_2(-z,-w)\right)\mathrm L_1(z,w)
\biggr]
&\notag\\&\qquad\qquad\qquad
+ 2\frac{w-3}{w^2} \mathrm L_2(z,w) -\frac 2w\mathrm L_1(z,w)
\Biggr\}.
\end{flalign}

Here,
\begin{align}
\widehat{D}_L(w) &=-\frac1{24}\ln^3 w_+ 
			\Big( \ln w_+ -4\ln w_- \Big)
 -  3 \Li_{13}\left(w_-\right)
 +  6 \Li_4\left( w_- \right)
\notag\\&\quad
 -\Big( \ln w_+ + 2 \ln w_- \Big)\Li_3\left(w_-\right)
- \frac12\zeta_2\ln^2 w_- + 2\zeta_3\ln w_-\,,
\end{align}
where $w_\pm = (1\pm w)/2$, and we use the following notation:
\begin{align}
\mathrm L_{\vec n}(z,w)=\begin{cases}
\ln\left(\frac{1-z}{1-w}\right) & \vec n=1
\\
\Li_{\vec{n}}\left(1,\dots,1,\frac {z - w}{1 - w}\right) &\vec{n}\neq 1
\end{cases}
\end{align}
and 
\begin{align}
\mathrm M_{22}(z,w)=\Li_{22}\left(\frac {z - w}{1 - w},\frac{w-1}{w+1}\right)\,,
&&
\widehat {\,\mathrm M}_{22}(z,w)=\Li_{22}\left(\frac{1+z}{1 + w },\frac{z + w}{1 + z}\right) \,,
\end{align}
where $\Li_{\vec{n}}(\vec{z})$ are the multiple polylogarithms defined in \eqref{eq:lidef}.
The hatted letters stand for the following combinations
\begin{align}
\widehat{L}_3(z,w) & =
	\mathrm L_3(z^2,w^2) - \mathrm L_3(z,w)  -\mathrm L_3(-z,w) -\mathrm  L_3(z,-w) -\mathrm  L_3(-z,-w),
\notag\\[2mm]
 \widehat{L}_{12}(z,w) & =
	\mathrm L_{12}(z^2,w^2) - \mathrm L_{12}(z,w)  -\mathrm L_{12}(-z,w) -\mathrm  L_{12}(z,-w) -\mathrm  L_{12}(-z,-w).
\end{align}
All $B_L$-functions are regular at $z=w$. Moreover, $B_L^{(2,\beta)}(w,w) =  B_L^{(2,P)}(w,w) =0$, and 
for the non-planar contribution the following identity holds: $B_L^{(2,NP)}(w,w) -B_L^{(2,NP)}(-w,-w) =0$. 
It can also be checked that $B^{(2,NP)}_L(z,w)$ is regular at the point $w=0$.
Thus the longitudinal CF is a real analytic function in the whole Euclidean domain $-1<z<1$, $-1<w<1$ with 
logarithmic branching cuts outside this region.

%%%%%%%%%%%%%%%%%%%%%%%%%%%%%%%%%%%%%%%%%%%%%%%%%%%%%%%%%%%%%%%%%%%%%%%%%%%%%%%%%%%%%%%%%%%%%%%%%%%%%%%%%%%%%%%%%%%%%%%%%%%%%%%%%%%%%%%

                          \subsection{Transverse  CF}\label{App:TwoLoopTrans}

%%%%%%%%%%%%%%%%%%%%%%%%%%%%%%%%%%%%%%%%%%%%%%%%%%%%%%%%%%%%%%%%%%%%%%%%%%%%%%%%%%%%%%%%%%%%%%%%%%%%%%%%%%%%%%%%%%%%%%%%%%%%%%%%%%%%%%%

The resulting expression for the transverse CF  can be brought to the form in~Eqs.~\eqref{loopCFs}, \eqref{color}, \eqref{RGlogs}.
The double logarithmic contributions $\sim \ln^2(Q^2/\mu^2)$ are rather simple in this case too:
\allowdisplaybreaks
\begin{align}
A^{(2,\beta)}_{\perp,\ln^2}(z,w)  & = -\mathrm L_1(z,w) -\frac34\,, 
\notag\\
B^{(2,\beta)}_{\perp,\ln^2}(z,w)  & = \frac12({1+w})\,\mathrm L_1(z,w)\,, 
\notag\\
A^{(2,P)}_{\perp,\ln^2}(z,w)  & = 4\mathrm L_2(z,w) +4 \mathrm L_1^2(z,w) -\mathrm L_1^2(w,-w)
                       + 6\mathrm L_1(z,w) + \frac94\,, 
\notag\\
B^{(2,P)}_{\perp,\ln^2}(z,w)  & = 
(1+w)\Big[ \mathrm L_2(w,-w) \!-\!\mathrm L_2(z,-w) \!-\!2\mathrm L_1(z,w)\Big(\mathrm L_1(z,-w) +\frac32\Big) 
                             -\mathrm L_1(-z,-w)\Big], 
\notag\\
A^{(2,NP)}_{\perp,\ln^2} (z,w)&= B^{(2,NP)}_{\perp,\ln^2}(z,w) =0\,.
\end{align}
For the single-logarithmic contributions $\sim \ln(Q^2/\mu^2)$ we obtain
\begin{align}
A^{(2,\beta)}_{\perp,\ln}(z,w) & =
2\mathrm L_2(z,w) - \ln^2(1-z) + \ln^2(1-w) + \frac{29}{6} \mathrm L_1(z,w) - \frac32\ln(1-w)  +\frac{19}4,
\notag\\
B^{(2,\beta)}_{\perp,\ln}(z,w)  & = 
\frac{1+w}{2}\Big[ \ln^2(1-z) -\ln^2(1-w)-2\mathrm L_2(z,w) -\frac {19}3\mathrm L_1(z,w) -\mathrm L_1(-z,-w)
\Big],
\notag\\
A^{(2,P)}_{\perp,\ln}(z,w)  & =   8\mathrm L_3(z,w) + 4\mathrm L_1^3(z,w)  
                  + 4\Big[2\ln(1-w) - 3\Big]\mathrm L_2(z,w)
                  \notag\\
                  &\quad
                  +\Big[4\ln(1-w^2)-3 \Big]\mathrm L_1^2(z,w)
                  - \mathrm L_1(z,w)\Big[4\mathrm L_1^2(w,-w) - 6\ln(1-w^2)  + \frac{103}{6} \Big]
                  \notag\\
                  &\quad
                  - \frac{47}{4} + \frac13\ln^3(1-w^2) -\frac{8}{3}\ln^3(1-w) + \frac92\ln(1-w),
\notag\\
B^{(2,P)}_{\perp,\ln}(z,w)  & =- 2(1 + w)\biggl\{ \mathrm L_{12}(z,w) + \mathrm L_3(z,-w) + \mathrm L_1^3(z,w)
                               - \frac32\mathrm L_2(z,w) - \frac32\mathrm L_2(-z,w)
                             \notag\\
                             &\quad
                             + \big(\ln(1+w)-1\big)\mathrm L_2(z,-w)   %- \mathrm L_2(z,-w)  
                               - \frac12\Big( \ln(1+w) -5\ln(1-w) + 3 \Big)\mathrm L_1^2(z,w)
                             \notag\\
                             &\quad
                               - \mathrm L_1^2(-z,-w)
                               - \frac12\mathrm L_1(z,w)\left(\ln^2(1-w^2) -4\ln^2(1-w)-6\ln(1+w)+\frac{65}{6}\right)
                             \notag\\
                             &\quad
                               - \frac12\mathrm L_1(-z,-w)\left(\ln(1+w)
                               -3\ln(1-w) +\frac{19}2\right)
                               %\mathrm L_1(-z,-w) %+ 3\ln(1+w)\mathrm L_1(z,w)
                             %
                               %+ \frac32\ln(1-w)\mathrm L_1(-z,-w) 
                               \notag\\
                             &\quad
                               - \mathrm L_3(w,-w) - \big(\ln(1+w) -1\big )\mathrm L_2(w,-w) 
                               + \frac32\mathrm L_2(-w,w) 
                         \biggr\},
 \notag\\
 A^{(2,NP)}_{\perp,\ln}(z,w)  & = - 2\widehat{L}_3(z^2,w^2) +4\widehat{L}_{12}(z,w) %+ 4 \Big( \mathrm L_3(z,w) + \mathrm L_3(-z,-w) \Big) 
                  + 8\mathrm L_1(z,w)\mathrm L_2(-z,-w)  + 2\mathrm L_1^2(w,-w)\mathrm L_1(z,w) 
\notag\\
&\quad
                  %+ 4\Big( \mathrm L_{12}(z^2,w^2) - 2\mathrm L_{12}(z,w) - 2\mathrm L_{12}(-z,-w) )
                 %
                 + \frac83\mathrm L_1(z,w) + \frac12,
\notag\\
B^{(2,NP)}_{\perp,\ln}(z,w)  & =   4w\Big( \mathrm P_{12}(z,w) +  \mathrm L_3(z,w) -\frac12 \widehat{L}_{12}(z,w)\Big)
				- 2(1-w)\mathrm L_1(w,-w)\mathrm L_2(z,-w)
\notag\\
&\quad
                 - 2(1+w)\mathrm L_1(z,w)\Big(\mathrm L_2(-z,-w) + \mathrm L_2(-z,w) + \frac12\mathrm L_1^2(w,-w) \Big) 
\notag\\
&\quad
                 -\left(\frac{10}3 -\frac23w\right)\mathrm L_1(z,w) 
                 - 2w\Big(\mathrm L_{12}(w,-w) + \mathrm L_{12}(-w,w)\Big)  
 \notag\\
&\quad                
                 + 2(1-w)\mathrm L_1(w,-w)\mathrm L_2(w,-w).                                    
\end{align}    
The function $\mathrm P_{12}$ is defined in Eq.~\eqref{defMP}.

The remaining contributions are:

\noindent $\bullet$~~Terms $\sim \beta_0 C_F$:
\begin{subequations}
\label{ABbeta2}
\begin{align}
\label{Aperpbeta}
A_\perp^{(2,\beta)}(z,w)  &=
\mathrm L_3(z,w) - 2\,\mathrm L_{12}(z,w) + 2\,\mathrm L_2(z,w)\left(\ln(1-w) -\frac {7}3\right)  -\frac34\zeta_2 -\frac{25}{48}
\notag\\
&\quad
    + \frac53\biggl(\ln^2(1-z) - \ln^2(1-w)\biggr)
              -\left(\frac{121}{36} + \zeta_2\right)\mathrm L_1(z,w) %\ln(1-z)
+\frac14\ln(1-w)
\notag\\
&\quad + \biggl\{ - \frac13 \bigl[ \ln^3 (1 - z) - \ln^3 (1 - w) \bigr] + \frac{3}{4} \bigl[\ln^2(1 - z) - 2\ln^2(1 - w) \bigr] 
 \notag\\ &\quad 
                       + \zeta_2\mathrm L_1(z,w)  - \frac{7}{ 2}\ln(1-z) + 8\ln(1-w) - 9 + \frac34\zeta_2
\biggr\},
\\[2mm]
\label{BperpBeta}
B_\perp^{(2,\beta)}(z,w) &=
\frac{1+w}2\biggl[ 2\, \mathrm L_{12}(z,w)
                              -\mathrm L_3(z,w) %\left(\frac{w-z}{1+w}\right)
                 -2\,\mathrm L_{2}(z,w) %\left(\frac{w-z}{1+w}\right)
                 \left[\ln(1-w)-\frac{10}3\right] +\frac{31}9\mathrm L_1(z,w)
\notag\\
&\quad                 
                 + \zeta_2\mathrm L_1(z,w)\biggr]
-\frac{w+4}{3}\bigl[\ln^2(1-z)-\ln^2(1-w)\bigr]
-2\mathrm L_{2}(z,w) +\frac16 \mathrm L_1(z,w) 
\notag\\
&\quad
+\biggl\{
(1+w)\biggl(\frac16\mathrm L_1^3(z,w) + \frac12\ln(1-w)\mathrm L_1^2(z,w)
                     - \frac34\biggl[\mathrm L_1^2(z,w) + \mathrm L_1^2(-z,-w)\biggr]\biggr)
\notag\\&\quad                     
                     - \mathrm L_1(z,w) \left( w\left[3\ln(1 - w)  - 6\right]  + \frac{1+w}2\zeta_2 - 2 - \frac12(1+w)\ln^2(1-w)\right)
\biggr\}.
\end{align}
\end{subequations}
The terms in the curly brackets in Eq.~\eqref{ABbeta2} originate from the $\epsilon$-correction term~$C^{(1,1)}$, see Eq.~\eqref{eq:d=4-restored}.

\noindent $\bullet$~~Planar contributions $\sim C_F^2$:
\begin{subequations}
\begin{align}
&A_\perp^{(2,P)}(z,w) =
\notag\\&=
4\biggl\{\frac{701}{192} +\frac92\zeta_3+ \mathrm L_4(z,w) +\mathrm L_{13}(z,w) +5\,\mathrm L_{112}(z,w)
    +\mathrm L_1(z,w)\Big(\mathrm L_3(z,w ) +2\mathrm L_{12}(z,w)\Big)
\notag\\
&\quad
   -\frac32 \mathrm L_{12}(z,w)  +\left(2\ln(1-w)-\frac 94\right)\mathrm L_3(z,w) +\frac14\left[\ln^2(1-z)-\ln^2(1-w)\right]^2
\notag\\
&\quad
    -\frac1{16}\left[\ln^2(1-w)-\ln^2(1+w)\right]^2 - \frac34\Big[\ln^2(1-z)-\ln^2(1-w)\Big]%\ln\left(\frac{1-z}{1-w^2}\right)
    \Big(\mathrm L_1(z,w)-\ln(1+w)\Big)
\notag\\
&\quad
+\left(\ln^2(1-w)- 3\ln(1-w) +\frac{37}{12}\right)\mathrm L_2(z,w)-\frac94\ln(1-w)\left[\ln(1-z)-\frac12\ln(1+w)\right]
\notag\\
&\quad
-\frac{49}{48}\ln^2(1-z) +\frac{65}{24}\ln^2(1-w)
+ \left(6\zeta_3+\frac{167}{144}\right) \mathrm L_{1}(z,w)
-\frac{47}{16}\ln(1-w)\biggr\}\,,% \hskip 10mm  \checkmark
\\[2mm]
&B_\perp^{(2,P)}(z,w) =
\notag\\&
= (1 + w)\biggl\{ - \mathrm L_1(z,w)\Big[\mathrm L_3(z,w) + 2\mathrm L_{12}(z,w) \Big]- 3\mathrm L_{13}(z,w)
                            - 3\mathrm L_{112}(z,w)  + \frac32 \mathrm L_3(z,w)
 \notag\\ &\quad
 - \ln(1-w)\Big [ 2\mathrm L_{12}(z,w) - 3\mathrm L_2(z,w)\Big ]  % + \frac32 \mathrm L_3(z,w)
 - \frac83\mathrm L_2(z,w)  - \mathrm L_4(z,-w)  - \mathrm L_{13}(z,-w)
\notag\\ &\quad
 - \mathrm L_1(z,-w) \Big [ \mathrm L_3(z,-w) %- \mathrm L_{13}(z,-w)
 + 2 \mathrm L_{12}(z,-w) \Big ] -5\mathrm L_{112}(z,-w)
                            - 2\ln(1+w)\mathrm L_3(z,-w)
\notag\\
 &\quad
                            - \ln^2(1+w)\mathrm L_2(z,-w) +  \mathrm L_{12}(z,-w)
                            - \frac12\Big[ \ln^2(1 - z) - \ln^2(1-w) \Big] \Big[
                            \ln^2(1 - z)
\notag\\
&\quad
                            - \ln^2(1 + w)  + 3\ln(1 - w^2) 
                            \Big ] - 12\zeta_3 \mathrm L_1(z,w)
                           + \frac12\mathrm L_3(-z,w)  - \mathrm L_1(-z,-w)\mathrm L_2(-z,-w)
 \notag\\
&\quad
                             - \mathrm L_1(-z,w)\mathrm L_2(-z,w) - 3\mathrm L_{12}(-z,w) 
                            + \frac52\Big[ \mathrm L_3(z,-w) + \mathrm L_3(-z,w) \Big]
 \notag\\
&\quad
                            + 2\Big[\ln^3(1-z)  -\ln^3(1-w)\Big]  + \frac23\big[\ln^3(1+z) -\ln^3(1+w)\Big] + 2\ln(1+w)\mathrm L_2(z,-w)
\notag\\
&\quad
                       + \Big(3\ln(1-w) -1\Big)\mathrm L_2(-z,w)
                             - \frac52\Big[( \mathrm L_2(z,-w) + \mathrm L_2(-z,w) + \frac12\mathrm L_1^2(w,-w)\Big]
\notag\\
&\quad
 -3\Big[ \mathrm L_2(z,w) + \mathrm L_2(-z,-w) \Big]   - \frac32 \left[\ln^2(1 - w) + \ln^2(1 + w) \right]
 \Big[\mathrm L_1(z,w) + \mathrm L_1(-z,-w) \Big]
  \notag\\
&\quad
                          +\frac12\ln(1+w)\mathrm L_1(z,w) +\frac12\ln(1-w)\mathrm L_1(-z,-w)
                             + \frac76\Big[ \ln^2(1-z) - \ln^2(1-w)\Big]
  \notag\\
&\quad
                             +\frac 32 \Big[\ln^2(1+z) - \ln^2(1+w)\Big]
                             + \ln(1 - w^2) \Big[4\mathrm L_1(z,w) + 3\mathrm L_1(-z,-w)\Big]
  \notag\\
&\quad
                             - \frac{295}{36}\mathrm L_1(z,w) - \frac{223}{12}\mathrm L_1(-z,-w)  - \widehat{D}_\perp(w) \biggr\},
\end{align}
\end{subequations}
where
\begin{align}
\widehat{D}_\perp(w) &= - \mathrm L_4(w,-w) - \mathrm L_1(w,-w) \mathrm L_3(w,-w) - \mathrm L_{13}(w,-w)
                      -2\mathrm L_1(w,-w) \mathrm L_{12}(w,-w)
\notag\\
&\quad
                       - 5 \mathrm L_{112}(w,-w)
              - 2\ln(1 + w)\mathrm L_3(w,-w) - \ln^2(1 + w) \mathrm L_2(w,-w) +  \mathrm L_{12}(w,-w)
 \notag\\
&\quad
              - \mathrm L_1(-w,w)\mathrm L_2(-w,w) - 3\mathrm L_{12}(-w,w)
              + \frac12\mathrm L_3(-w,w) + 3\ln(1 - w) \mathrm L_2(-w,w)  &
\notag\\
&\quad
            - \mathrm L_2(-w,w)   + \frac52( \mathrm L_3(w,-w) + \mathrm L_3(-w,w) )  +  2\ln(1 + w)\mathrm L_2(w,-w).
\end{align}
It can be checked that $B_\perp^{(2,P)}(w,w)=0$.

\noindent $\bullet$~~Non-planar contributions $\sim C_F/N_c$:
\begin{subequations}
\begin{align}
&A_\perp^{(2,NP)}(z,w) =
\notag\\& =
  2\mathrm L_{13}(z^2,w^2)
                         - 4\mathrm L_1(z,w)\mathrm L_3(z^2,w^2)
                         - 6 \mathrm L_2(z,w) \mathrm L_2(-z,w)
                         + 12\mathrm L_1(z,w)\mathrm L_3(-z,-w)
\notag\\
&\quad
                    -\mathrm L_4(z^2,w^2) -4\mathrm L_2(z,w)\mathrm L_2(z,-w)
                    - 4\Big[ \widehat{\, \mathrm M}_{22}(-z,w) - \widehat{\,\mathrm M}_{22}(w,-z) - \frac12 \mathrm L^2_1(z,w)\mathrm L_2(-z,w)\Big]
\notag\\
&\quad
+2\Big[ \widehat{\, \mathrm M}_{22}(z,w) - \widehat{\,\mathrm M}_{22}(w,z) - 4\mathrm L_1(z,w)\mathrm L_{12}(-z,w)
	+ 2\mathrm L_1(-z,w)  \widehat{L}_{12}(z,w) \Big]
\notag\\
&\quad	
 	+\frac32 \mathrm L_1^2(-z,w) \Big[\mathrm L_2(z,w) +\mathrm L_2(z,-w)\Big]
	 +2\mathrm L_2^2 (z,w) + 8\mathrm L_1(z,w)\mathrm L_3(z,w) -16\mathrm L_{13}(z,w)
\notag\\
&\quad
 - 2\mathrm L_1^2(z,w)\mathrm L_2(z,w) - 4\mathrm L_1(z,w)\mathrm L_{12}(z,w)	+  2\Big[ \mathrm L_4(z,w) + \mathrm L_4(-z,w)\Big]
 -4\mathrm L_{13}(-z,w)
\notag\\
&\quad
+ 2\mathrm L_1^2(-z,w)\mathrm L_2(-z,w) + 8\mathrm L_1(-z,w)\mathrm L_{12}(-z,w) + 12\mathrm L_{112}(-z,w)
\notag\\
&\quad
-\ln(1-w^2)\Big[ \mathrm L_3(z^2,w^2) - 2\mathrm L_{12}(z^2,w^2) + 4 \mathrm L_{12}(z,w)\Big]
 + 3\mathrm L_1(w,-w)\mathrm L_1(-z,w)\mathrm L_2(z,w)
\notag\\
&\quad
-2\Big[\ln(1+w)-3\ln(1 - w)\Big]\Big[2\mathrm L_1(z,w)\mathrm L_2(-z,w) - 2\mathrm L_{12}(z,-w) +\mathrm L_3(z,w)
\notag\\
&\quad
+\mathrm L_3(-z,-w)\Big] - \left[\frac32 \mathrm L_1^2(w,-w)+\frac73\right]\mathrm L_2(z,w) +\frac43\mathrm L_1^2(z,w)
\notag\\
&\quad
+\Big[ \mathrm L_1(w,-w)\Big(7\ln(1-w)+\ln(1+w)\Big)+4\Big]\mathrm L_2(-z,w)
 +\mathrm L_1^2(w,-w)\mathrm L_1^2(-z,w)
 \notag\\
 &\quad
  -\left[\frac13\ln^3(1-w^2) -\frac43\ln^3(1-w) -\frac43\ln^3(1+w) -\frac43\ln(1-w^2)+\frac{ 73}{18} -24\zeta_3\right]\mathrm L_1(z,w)
\notag\\
&\quad
+\frac14\ln(1-w^2) +\frac13\mathrm L_1^2(w,-w) - \frac{13}{24} \mathrm L^4_1(w,-w) + 27\zeta_3 - \frac{73}{24}\,,
\\[10mm]
&B_\perp^{(2,NP)}(z,w)=
\notag\\ & =
	(w + 3)\mathrm L_1(z,w) \mathrm L_3(z^2,w^2)
      + (2 w + 3) \mathrm L_2(z,w)\mathrm L_2(-z,w)
     -w \Big[ \widehat{\,\mathrm M}_{22}(z,w) -\widehat{\,\mathrm M}_{22}(z,-w)
\notag\\
&\quad  + \mathrm L_{13}(z^2,w^2) \Big]
        + 2(2-w)\Big[\mathrm L_1(-z,-w)\mathrm L_3(z,w) + \mathrm L_1(-z,w)\mathrm L_3(z,-w)\Big]
        +\mathrm M_{22}(z,w)
\notag\\
&\quad
+ 7 \mathrm M_{22}(z,-w)
+3\mathrm L_2(z,w)\mathrm L_2(z,-w)
+2w\Big(\mathrm P_{211}(z,-w) +2\mathrm P_{121}(z,w)
+\mathrm P_{13}(z,w)
\notag\\
&\quad
+ \mathrm P_{13}(z,-w)  + \mathrm P_{31}(z,-w)
- \mathrm P_{22}(z,w) + \mathrm P_{22}(z,-w)
\Big)
\notag\\
&\quad
-2(w+3)\Big[ \mathrm M_{13}(z,w)\! +\! \mathrm M_{13}(z,-w)\! +\!  \mathrm M_{112}(z,w)\! +\! \mathrm M_{112}(z,-w)
\!+\!\mathrm M_{22}(z,w)\! +\!\mathrm M_{22}(z,-w)
\Big]
\notag\\
&\quad
-2(1+w)\Big[\mathrm M_{121}(z,w) +\mathrm M_{121}(z,-w) \Big] - (2w-5)\mathrm L_4(z,w) + (4w-1) \mathrm L_4(z,-w)
\notag\\
&\quad
+(9 w+7) \mathrm L_{13}(z,w) + (5 w+7)\mathrm L_{13}(z,-w) + 2\mathrm L_{22}(z,w) +2(1+w)\mathrm L_{22}(z,-w)
\notag\\
&\quad
+(w+3)\Big[\mathrm L_{31}(z,w) +\mathrm L_{31}(z,-w) \Big] + \mathrm L_{121}(z,w)  + (2w+1)\mathrm L_{121}(z,-w)
\notag\\
&\quad
-(3w + 1)\mathrm L_{211}(z,w) + (w-1)\mathrm L_{211}(z,-w) +w \Big[\ln(1-w^2) \mathrm L_3(z^2,w^2) -10 \mathrm   P_{12}(z,w)
\notag\\
&\quad
-4\ln(1-w)\mathrm   P_{12}(z,-w) \Big]  - \Big[2(1+w)\ln(1 + w) + 5(1 - w) \Big] \mathrm M_{21}(z,w)
\notag\\
&\quad
 - \Big[2(1+w)\ln(1 - w) + 5(1 - w) \Big] \mathrm M_{21}(z,-w)  + w\mathrm L_1(w,-w)\Big[\mathrm L_{21}(z,w) +\mathrm L_{21}(z,-w)\Big]
\notag\\
&\quad
-\Big[2w\ln(1-w^2) - 2(1 - w)\ln(1 - w) - 5(1 - w) \Big]\mathrm L_3(z,w)
\notag\\
&\quad
+\Big[6(1 - w)\ln(1 - w) - 4(1+w)\ln(1 + w)   + 5(1 + w) \Big]\mathrm L_3(z,-w)
\notag\\
&\quad
+ (w+2)\mathrm L_1^2(-w,w)\Big [\mathrm L_2(z,w) +\mathrm L_2(z,-w)\Big]
- \Big[\frac12 \ln^2w_- + \Li_2(w_+) - \frac{10}3 w_+ - \zeta_2\Big]\mathrm L_2(z,w)
\notag\\
&\quad
-\Big[\frac72\ln^2 w_+ + 7 \Li_2(w_-) - w -7\zeta_2\Big]\mathrm L_2(z,-w)  + \mathrm L_1^2(z,w)\Big[\zeta_2(w + 3)  + \frac{10}3 w - \frac{25}6\Big]
\notag\\
&\quad
+\mathrm L_1(z,w)\Big[ \frac13(2w+3)\Big(\ln^3w_- +\ln^3 w_+\Big) -(w+2)\ln w_+\ln w_-\ln(w_+w_-)
\notag\\
&\quad
                          - (1+w)\mathrm L_1(w,-w) \Big(\Li_2(w_-) - \Li_2(w_+) \Big)
                          - 2(3+w) \Big(\Li_3(w_+) + \Li_3(w_-) \Big)
 \notag\\
&\quad
                          +\zeta_2\Big((1+w) \ln(1-w^2) + 2(w+3) \ln w_-  + 5(1-w)\Big)
  \notag\\
&\quad
                          -\frac52 (1-w) \mathrm L_1(w,-w) - \frac 2 3 (5-w)\ln(1-w) -\frac19 (8 w -49)
                         \Big]  - \widehat{D}^{(NP)}_\perp(w)
\notag\\
&\quad
- \frac{3(w^2-1)}{w^2} \Big( \mathrm L_1(-z,-w)\mathrm L_2(z,w) + \mathrm L_1(-z,w)\mathrm L_2(z,-w)  - 2w\mathrm L_2(z,-w) 
                       \notag\\
&\quad
                                        - \frac14\mathrm L_1(z,-z)\mathrm L_1^2(w,-w) -\frac12 w\mathrm L^2_1(w,-w) 
                                        - \frac{w(w+2)}{w+1}\mathrm L_1(z,w) \Big)     
\notag\\
&\quad
 +(w-z)\Biggl(  \frac{ 6(2w+3)(w-1)}{w^2}\mathrm L_2(z,w)
                                  + \frac{9}{w^4}(w^2-1)\biggl( 
                                   z\biggl[ 
                                         \frac12\mathrm L_2(-z,w)\mathrm L_2(z,w)  +  \widetilde{\mathrm M}_{22}(z,w) 
\notag\\
&\quad                                         
                                          + \frac12\mathrm L_1^2(w,-w)\mathrm L_2(z,w) 
             - \frac14\mathrm L_1(w,-w)\bigl( \widehat{ L}_3(z,w) -6\zeta_3
             %ML3(z^2,w^2) - ML3(z,w)  - ML3(-z,w) - ML3(z,-w) - ML3(-z,-w) - 6*zeta[3] 
             \bigr) 
                               - \frac1{24}\big( \ln^4 w_+ - 4\ln^3w_+\ln w_-\big) 
\notag\\
&\quad
                               - (\ln w_+ +  2\ln w_-) \mathrm L_3(-w,-1) + 6\mathrm L_4(-w,-1)
                                        - 3\mathrm L_{13}(-w,-1)  - \frac12\zeta_2\ln^2 w_- - 4\zeta_3\ln w_-
                                       \biggr]
  \notag\\
&\quad                               
                                  + 2w\left[ \mathrm L_1(-z,-w)\mathrm L_2(z,w) - \frac18\mathrm L^2_1(w,-w)\mathrm L_1(z,-z) \right] \biggr)
                                - 6z\frac{(2w^2-3)}{w^3}  \biggl[  \frac14\big(\widehat{ L}_3(z,w) -6\zeta_3\big)
\notag\\
&\quad
                                         - \mathrm L_1(-z,-w)\mathrm L_2(z,w) -\frac12  \widehat{ L}_{12}(z,w)
                                         -\frac14\mathrm L_1^2(w,-w)\mathrm L_1(z,w)
              \biggr]
                                   - \frac{3}{w}\mathrm L_1(z,-z)
                                        \Biggr)\,.                                                             
\end{align}
\end{subequations}

The subtraction term  $\widehat{D}^{(NP)}_\perp(w)$ is determined by condition $B_\perp^{(2,NP)}(w,w) =0$ and takes the form
\begin{align}
\widehat{D}^{(NP)}_\perp(w) & =  - w\widehat {\, \mathrm M}_{22}(w,w)
                       + 2(2-w)  \mathrm L_1(-w,w)\mathrm L_3
                      + 7\mathrm M_{22}
                      + 2w(\mathrm P_{211}  + \mathrm P_{13}
                      + \mathrm P_{31}   +  \mathrm P_{22})
 \notag\\
 &\quad
                      - 2(w+3) ( \mathrm M_{13} + \mathrm M_{22} + \mathrm M_{112})
                      - 2(1+w)\mathrm M_{121}
                      + (4w - 1)\mathrm L_4
                      + (5w+7)\mathrm L_{13}
                      \notag\\
 &\quad
                      + 2(1+w)\mathrm L_{22}
                      + (w + 3)\mathrm L_{31}
                      + (2w+1)\mathrm L_{121}
                      - (1 - w)\mathrm L_{211}  - 4w\ln(1 - w)\mathrm P_{12}
\notag\\ &\quad
                      - ( 2(1 + w)\ln(1 - w) + 5(1 - w) )\mathrm M_{21}    + w\mathrm L_1\mathrm L_{21}   + (w+2)\mathrm L^2_1\mathrm L_2
\notag\\
&\quad
                      + ( - 4(1+w)\ln(1 + w) + 6(1 - w)\ln(1 - w) + 5(1 + w) )\mathrm L_3
\notag\\&\quad
                      - \left(\frac72 \ln^2 w_+   + 7\Li_2(w_-) - w -7\zeta_2\right)\mathrm L_2,
\end{align}
where we used a shorthand notation $(\mathrm L,\mathrm M,\mathrm P)_{a,b,c}\equiv (\mathrm L,\mathrm M,\mathrm P)_{a,b,c}(w,-w)$, and
\begin{align}\label{defMP}
\mathrm P_{\vec{n}}(z,w) &=\Li_{\vec{n}}\left(1,\ldots,1,\frac{w-z}{2w},\frac{2w}{w-1}\right)
\notag\\
\mathrm M_{n_1,n_2}(z,w) &=\Li_{n_1,n_2}\left(\frac {z - w}{1 - w},\frac{w-1}{w+1}\right), 
\notag\\
\mathrm M_{112}(z,w) &=\Li_{112}\left(\frac {z - w}{1 - w},1,\frac{w-1}{w+1}\right),
\notag\\
\mathrm M_{121}(z,w) &=\Li_{121}\left(1,\frac {z - w}{1 - w},\frac{w-1}{w+1}\right)
\end{align}
and
$$
\widetilde{\mathrm M}_{2,2}(z,w) =\Li_{22}\left(\frac{w-1}{w+1},\frac {z - w}{1 - w}\right). 
$$

%%%%%%%%%%%%%%%%%%%%%%%%%%%%%%%%%%%%%%%%%%%%%%%%%%%%%%%%%%%%%%%%%%%%%%%%%%%%%%%%%%%%%%%%%%%%%%%%%%%%%%%%%%%%%%%%%%%%%%%%%%%%%%%%%%%%%%%
\section{Threshold expansion}\label{App:ThresholdExpansion}
%%%%%%%%%%%%%%%%%%%%%%%%%%%%%%%%%%%%%%%%%%%%%%%%%%%%%%%%%%%%%%%%%%%%%%%%%%%%%%%%%%%%%%%%%%%%%%%%%%%%%%%%%%%%%%%%%%%%%%%%%%%%%%%%%%%%%%%

In this Appendix, we provide threshold expansions of our results for the coefficient functions.
The transverse CF is singular at the points $z=\pm 1$
\begin{align}
C_{\perp}^{(k)}\left(z,\omega,\frac{Q^2}{\mu^2}\right) &=\frac{w C_F}{1-z}\Big[
 A_\perp^{(k)}\left(z,\omega,\frac{Q^2}{\mu^2}\right) + A_\perp^{(k)}\left(z,-\omega,\frac{Q^2}{\mu^2}\right)
\Big] +\ldots
\end{align}
and the functions $A_\perp^{(2)}\left(z,\omega,\frac{Q^2}{\mu^2}\right)$ contain a series of logarithmic contributions 
$\sim \ln \bar z$, $\bar z = 1-z$ up to power $\ln^{2k}\bar z$. In what follows we collect the corresponding expressions for the sum
\begin{align}
\widetilde A_\perp^{(k)}\left(z,\omega,\frac{Q^2}{\mu^2}\right)  & =   A_\perp^{(k)}\left(z,\omega,\frac{Q^2}{\mu^2}\right) + A_\perp^{(k)}\left(z,-\omega,\frac{Q^2}{\mu^2}\right).
\end{align}
At one loop, up to terms $\mathcal{O}(\bar z^1)$, one gets
\begin{align}
\widetilde  A^{(1)}_\perp(z,w) &=
2 \ln^2\bar z - 3  \ln\bar z  -\ln^2(1+w)- \ln^2(1-w) +3 \ln(1-w^2)-9  + \ldots,
\notag\\
\widetilde A^{(1)}_{\perp,\ln}(z,w) &= 4 \ln\bar z -2\ln(1-w^2)+3 +\ldots. 
\end{align}
To the two-loop accuracy, we obtain
\begin{align}
\widetilde  A^{(2,\beta)}_\perp(z,w) & =   -\frac23 \ln^3\bar z 
+ \frac{29}{6} \ln^2\bar z  -\frac{247}{18} \ln\bar z   
+ \Big(2\zeta_2+\frac{209}{18}\Big) \,\ln(1-w^2)
-\frac{19}{6}\ln^2(1+w)
\notag\\[1mm]&\quad
-\frac{19}{6} \ln^2(1-w) +\frac13\ln^3(1+w)+\frac13\ln^3(1-w)
%\notag\\&\quad
- \frac{28}{3}\zeta_2-2\zeta_3 -\frac{457}{24} +\ldots,
\notag\\
\widetilde
 A^{(2,\beta)}_{\perp,\ln}(z,w) & =   
 -2 \ln^2\bar z + \frac{29}{3}\ln\bar z   
-\frac{19}{3}\ln(1\!-\!w^2)+\ln^2(1\!+\!w)+\ln^2(1\!-\!w)+4\zeta_2+\frac{19}{2}+\ldots\,,
\notag\\[1mm]
\widetilde
 A^{(2,\beta)}_{\perp,\ln^2}(z,w) & =   -2\ln\bar z +\ln(1-w^2) -\frac32\,,
\end{align}
%%%%%%%%%%%%%%%%%%%%%%%%%%
\begin{align}
\widetilde
 A^{(2,P)}_\perp(z,w) &=  2 \ln^4\bar z   -6 \ln^3\bar z
+ 
\biggl[6\ln(1-w^2)-2\ln^2(1+w)-2\ln^2(1-w)-\frac{49}{6}\biggr]\ln^2\bar z
\notag\\&\quad
+ \biggl[
3\ln^2(1+w)+3\ln^2(1-w) -9\ln(1-w^2) +72\zeta_3+\frac{167}{18}\biggr]\ln\bar z
\notag\\&\quad
%+\frac12 \Big(\ln^4(1+w)+\ln^4(1-w)\Big) -3\Big(\ln^3(1+w)+\ln^3(1-w)\Big)
+\frac12\Big(\ln^2(1-w)+\ln^2(1+w)-3\ln(1-w^2)\Big)^2
\notag\\&\quad
+ \Big(%\frac{65}{6}
\frac {19}{3} + 4 \zeta_2\Big)\Big[\ln^2(1+w)+\ln^2(1-w)\Big]
-\Big(28\zeta_3 + 12\zeta_2 + \frac{295}{18}\Big)\ln(1-w^2)
\notag\\[1mm]&\quad
+20 \zeta_2^2 + 30 \zeta_3+\frac{74}{3}\zeta_2 +\frac{701}{24}+\ldots\,,
\notag\\[2mm]
\widetilde
 A^{(2,P)}_{\perp,\ln}(z,w) &=  8 \ln^3\bar z
- 4\Big[\ln(1-w^2)+\frac32\Big] \ln^2\bar z
-4  \Big[\ln^2(1+w)+\ln^2(1-w)-\frac92 \ln(1-w^2)
\notag\\&\quad
+\frac{103}{12} \Big] \ln\bar z
+2\ln(1-w^2)\Big[\ln^2(1-w) + \ln^2(1+w) \Big]  %-12
+6\ln(1-w)\ln(1+w)
\notag\\&\quad 
-9\ln^2(1-w^2)
+\Big(\frac{65}{3}+8 \zeta_2\Big) \ln(1-w^2)+16\zeta_3
-24\zeta_2-\frac{47}{2} \ldots\,,
\notag\\[2mm]
\widetilde
 A^{(2,P)}_{\perp,\ln^2}(z,w) &= 
 8 \ln^2\bar z + 8\Big[\frac32 -\ln(1\!-\!w^2)\Big] \ln\bar z
+ 2\ln^2(1\!-\!w^2) -6\ln(1\!-\!w^2)+ 8\zeta_2+ \frac92 +\ldots\,,
\end{align}

\begin{align}
\widetilde
 A^{(2,NP)}_\perp(z,w) &= 
 \left(-4 \zeta_2 +\frac{8}{3}\right)\ln^2\bar z
 +  \left(40\zeta_3 -\frac{73}{9}\right)\ln\bar z 
+\Big(2\zeta_2-\frac43\Big)\Big[\ln^2(1+w) + \ln^2(1-w)\Big] 
\notag\\&\quad
+\Big(-26 \zeta_3 +\frac{41}{9}\Big)\ln(1-w^2)
+\frac{31}{5}\zeta_2^2-\frac{26}{3}\zeta_2 + 54\zeta_3 -\frac{73}{12}+\ldots\,,
\notag\\
\widetilde
 A^{(2,NP)}_{\perp,\ln}(z,w) &= 
\Big(-8\zeta_2+ \frac{16}{3}\Big) \ln\bar z + \Big( 4\zeta_2 -\frac83\Big)\ln(1-w^2) -12\zeta_3 +1+\ldots.
\end{align}
Note that the limits $\omega\to 1$ and $z\to 1$ do not commute so that the above expressions do not reduce to the corresponding DVCS results~\cite{Braun:2020yib,Schoenleber:2022myb} in the limit $\omega \to 1$.

%%%%%%%%%%%%%%%%%%%%%%%%%%%%%%%%%%%%%%%%%%%%%%%%%%%%%%%%%%%%%%%%%%%%%%%%%%%%%%%%%%%%%%%%%%%%%%%%%%%%%%%%%%%%%%%%%%%%%%
%%%%%%%%%%%%%%%%%%%%%%%%%%%%%%%%%%%%%%%%%%%%%%%%%%%%%%%%%%%%%%%%%%%%%%%%%%%%%%%%%%%%%%%%%%%%%%%%%%%%%%%%%%%%%%%%%%%%%%
%%%%%%%%%%%%%%%%%%%%%%%%%%%%%%%%%%%%%%%%%%%%%%%%%%%%%%%%%%%%%%%%%%%%%%%%%%%%%%%%%%%%%%%%%%%%%%%%%%%%%%%%%%%%%%%%%%%%%%

%\bibliography{references}

%\bibliographystyle{JHEP}

\providecommand{\href}[2]{#2}\begingroup\raggedright\endgroup

\end{document}